\documentclass[11pt]{article}
\usepackage[margin=1in]{geometry}

\usepackage{cite}
\usepackage{amsmath,amssymb,amsfonts}
\usepackage{graphicx}
\usepackage{textcomp}
\usepackage{xcolor}
\usepackage{booktabs}
\usepackage{array}
\usepackage{multirow}
\usepackage{caption}
\usepackage{adjustbox}  
\usepackage{placeins}  
\usepackage{float}
\usepackage{listings}
\usepackage{url}
\usepackage[hidelinks,breaklinks=true]{hyperref}
\usepackage{xurl}  

\usepackage{newunicodechar}
\newunicodechar{≈}{$\approx$}
\newunicodechar{≤}{$\leq$}
\newunicodechar{≥}{$\geq$}
\newunicodechar{∈}{$\in$}
\newunicodechar{∅}{$\emptyset$}
\newunicodechar{∪}{$\cup$}
\newunicodechar{∩}{$\cap$}
\newunicodechar{→}{$\rightarrow$}
\newunicodechar{←}{$\leftarrow$}
\newunicodechar{−}{$-$}
\newunicodechar{⇒}{$\Rightarrow$}
\newunicodechar{⊆}{$\subseteq$}
\newunicodechar{⊂}{$\subset$}
\newunicodechar{κ}{$\kappa$}
\newunicodechar{★}{$\bigstar$}
\newunicodechar{✓}{$\checkmark$}
\newunicodechar{✗}{$\times$}
\newunicodechar{§}{\S}
\newunicodechar{×}{\ensuremath{\times}}
\newunicodechar{·}{\ensuremath{\cdot}}

\providecommand{\tightlist}{%
  \setlength{\itemsep}{0pt}\setlength{\parskip}{0pt}}
\providecommand{\pandocbounded}[1]{#1}
\providecommand{\passthrough}[1]{#1}

\graphicspath{%
  {figures/}%
  {./}%
}

\makeatletter

\providecommand{\corresp}[1]{}
\providecommand{\history}[1]{}
\providecommand{\doi}[1]{}
\providecommand{\EOD}{}
\providecommand{\address}[2][]{}
\providecommand{\appendices}{\appendix}
\makeatother

\begin{document}

\title{Secrets That Survive Everything: Runtime Credential Exposure in Production Web Applications}

\author{Hemanth Gorijala\thanks{Independent Security Researcher, ORCID 0009-0006-9810-4001. E-mail: gorijala2k16@gmail.com. Published in \emph{IEEE Access}, 2026. DOI: 10.1109/ACCESS.2026.3734984. Open access under CC BY 4.0. An earlier version was posted as a preprint at Zenodo (DOI: 10.5281/zenodo.19464446); this manuscript is substantially extended, adding the GT-194 benchmark and the nine-scanner evaluation.}}
\date{}
\maketitle

\begin{abstract}
Pre-deployment secret scanning operates only on source code, never on what a production application serves. We document two exploitation chains in which Azure AD client credentials and APIM subscription keys from production JavaScript bundles enabled account takeover and mass data exposure. An authorized engagement covered approximately 2,000 enterprise web assets in one organization; 113 (5.65\%) served live credentials. To quantify the shift-right gap, we built an independent Ground Truth (GT-194) of 194 secret-grade credentials through Claude Opus 4.7 extraction and manual analyst review, with the 247 LLM-extracted candidates independently validated by GPT-5.5 (Brennan-Prediger $\kappa$ = 0.676). The principal finding is structural: 13.9\% of GT-194 (27 of 194) is surfaced only by manual analysis and recovered by none of the nine evaluated production scanners, a tool-agnostic blind spot the ground-truth model also misses. CryptoJS encrypted configuration separately defeats every static scanner: the credential exists only after decryption with a co-located key, reached only by runtime-aware detection. Combined coverage plateaus at 86.1\%. Among the nine scanners, the best static scanner recovers 36.6\% and the best runtime-aware scanner 77.8\% (F1 = 0.818, McNemar p \textless{} 0.001); the ground-truth model is reported separately as a reference comparator, not an evaluated detector. On 63 of 86 secret-exposed applications (73.3\%), the full Azure AD token-mint chain is co-located in one bundle, reachable from browser code. We characterize five paths by which credentials reach production undetected and present a layered runtime detection methodology and remediation framework. Recall is scoped to a single-organization Azure-heavy corpus (Section 10).
\end{abstract}

\medskip

\section{Introduction}\label{introduction}

Despite mature shift-left tooling for secret detection \cite{ref8, ref16}, live credentials continue to appear in the production-served content of enterprise web applications. Static secret scanners (GitLeaks, TruffleHog, GitHub Advanced Security) and SAST tools (Semgrep, SonarQube, Checkmarx) operate on source code repositories and build configurations and do not examine what a deployed application serves to its users. Credentials that reach production through build-time substitution, CI/CD pipeline variable injection, runtime-fetched configuration, third-party script inclusion, or deliberate scanner suppression therefore enter served content without crossing any layer those tools cover. This paper characterizes the resulting exploitation surface and quantifies the corresponding shift-right tooling gap. Prior population-scale work established that browser-delivered credential exposure exists; the contribution here is the complementary per-tool recovery benchmark and runtime-detection analysis for enterprise JavaScript bundles, not the existence of the phenomenon. Stated precisely, the specific problem addressed is: credentials that reach the production-served layer are not examined by any repository-stage detector, and the extent to which detectors operating on served content recover them has not been measured. We therefore ask three questions on a real enterprise corpus: (1) by what structural paths do live credentials reach served content without crossing a repository-stage scanner; (2) how much of a served-content credential ground truth do current static, template-driven, and runtime-aware detectors actually recover; and (3) how weaponizable are the credentials that do reach served content, measured by end-to-end exploitation chains.

An authorized security engagement covered approximately 2,000 enterprise web application assets across one organization. Of these, 113 (5.65\%, 95\% Clopper-Pearson exact CI: 4.7\%--6.8\%) contained at least one live credential in served content (JavaScript bundles, HTML view-source, or JSON and XML API responses) that had bypassed every pre-deployment secret scanning control in place. Of those 113 vulnerable applications, 63 (55.8\%, 95\% Clopper-Pearson exact CI: 46.1\%--65.1\%) contained a complete Azure AD credential set (client ID, client secret, tenant ID, and resource URI co-located in the same JS bundle) sufficient to execute the full exploitation chain documented in this paper. Not every credential exposure produces an exploitable end-to-end chain. A complete chain requires both the full credential set and over-permissive service principal scopes, as documented in Sections 3.4 and 4.3. Applications were selected based on engagement scope, not screened for likelihood of exposure. Confidence intervals are Clopper-Pearson exact intervals at the 95\% level.

To measure the shift-right tooling gap on the same corpus, we constructed an independent Ground Truth (GT-194) of 194 unique secret-grade credentials extracted from production JavaScript bundles across 113 enterprise applications. The benchmark is anchored to extraction by Claude Opus 4.7 (Anthropic) with no access to any of the evaluated tools, and its 247 LLM-extracted candidates were independently validated by an LLM from a different vendor (GPT-5.5, OpenAI; 207 concordant, Brennan-Prediger κ = 0.676), with the 27 manual-only additions verified by analyst review. After human review, manual analyst additions, and the field-deployment register of 8 May 2026, 194 unique secret-grade credentials were locked as the benchmark across eight credential types (Azure APIM subscription key, Azure AD client\_secret, CryptoJS-AES blob, plaintext user credential, JWT, CyberArk AIM, Google API key, and Other API key/token including App Insights iKey). Nine production scanners were evaluated against GT-194 at their tightened configurations, with Claude Opus 4.7 reported separately as an LLM-assisted ground-truth reference comparator (it co-constructed the benchmark and therefore cannot be scored as an independent detector). The evaluation uses pairwise McNemar significance testing with Holm-Bonferroni correction, an UpSet-style detection-set overlap analysis, and a separately tracked set of 249 chain-completion identifiers that are public-by-design and therefore not counted in the recall denominator. Section 8.5 reports the full results.

The remainder of this paper is organized as follows. Sections 3 and 4 document two end-to-end exploitation chains drawn from the corpus. Section 5 characterizes the five structural paths by which credentials reach production undetected. Sections 6 and 7 analyze why both shift-left and shift-right tooling categories systematically miss these paths. Section 8 presents a layered runtime detection methodology and the GT-194 evaluation. Sections 9 through 12 cover recommendations, limitations, future work, and responsible disclosure.

\subsection{Scope and Authorization}\label{scope-and-authorization}

Both exploitation chains described in this paper were identified during authorized security assessments. Organization names and application identifiers have been anonymized throughout.

\subsection{Contributions}\label{contributions}

This paper makes the following contributions:

\begin{enumerate}
\def\labelenumi{\arabic{enumi}.}
\item
  \textbf{Two documented exploitation chains}, end-to-end, reproducible attack paths in which Azure Active Directory client credentials and API subscription keys found in production JavaScript bundles were combined with over-permissive service principal scopes to achieve full account takeover and mass user data exposure in real enterprise applications that had passed all deployed shift-left security controls.
\item
  \textbf{A structural taxonomy of five paths to production} (build-time environment injection, CI/CD pipeline variable substitution, runtime configuration fetching, third-party script inclusion, and scanner suppression at the organizational level), each of which produces live credentials in deployed artifacts without those credentials appearing in the application owner's repository at any point.
\item
  \textbf{A characterization of the tooling gap}. A systematic analysis of why both shift-left scanning tools (GitLeaks, TruffleHog, SAST) and shift-right tools (DAST, WAF, RASP) fail to detect this class of vulnerability, grounded in the specific technical properties of minified production JavaScript.
\item
  \textbf{A detection methodology for the runtime layer}. A layered approach combining anchored vendor token patterns, Shannon entropy analysis, and key-value context scanning, designed for the constraints of production JavaScript: minification, noise suppression, and no outbound verification calls.
\item
  \textbf{A prioritized remediation framework}, covering immediate credential rotation, architectural remediation via the Backend for Frontend pattern, service principal scope correction, and ongoing monitoring controls for Azure AD and Azure API Management environments, with generalizations to AWS and GCP stacks.
\item
  \textbf{A cross-vendor LLM-validated benchmark (GT-194) and a quantified tool-agnostic blind spot.} Nine production scanners are evaluated against a locked benchmark assembled from independent Claude Opus 4.7 extraction and manual analyst review, with its LLM-extracted candidates independently validated by GPT-5.5 (Brennan-Prediger κ = 0.676, 207 of 247 concordant); Claude Opus is reported separately as an LLM-assisted ground-truth reference comparator rather than as an evaluated detector, since it co-constructed the benchmark. The evaluation uses pairwise McNemar significance testing with Holm-Bonferroni correction and UpSet-style detection-set overlap analysis. The principal finding is structural: 13.9\% of GT-194 (27 credentials) is recovered by none of the nine production scanners and is surfaced only by manual analysis. Separately, the CryptoJS encrypted-configuration class defeats every static scanner (the credential exists only after decryption with a co-located key) but is recovered by runtime-aware detection. Methodological mitigations of the developer-author conflict of interest are detailed in §8.5; recall figures are scoped to this corpus rather than cross-industry estimates (Section 10).
\end{enumerate}

\subsection{Related Work}\label{related-work}

\textbf{Secret leakage in version control.} The leakage of credentials through public repositories is well-documented. Meli et al.~\cite{ref8} conducted a large-scale measurement study of public GitHub repositories, identifying thousands of exposed API keys, passwords, and tokens across millions of commits, and established that secret leakage in source code is a systemic problem providing the empirical foundation for tools such as GitLeaks and TruffleHog. Demir et al.~\cite{ref15} extended this analysis to live web content, crawling 10 million web pages and identifying 1,748 distinct credentials from 14 cloud and SaaS providers embedded in JavaScript, HTML, and JSON resources, finding that 84\% of credentials appeared in JavaScript files and persisted for an average of twelve months. The present paper addresses a complementary structural problem: credentials that reach production without ever appearing in any repository, through build-time injection, pipeline variable substitution, and runtime configuration delivery.

\textbf{Industry-scale measurement of credentials in served JavaScript.} Intruder's December 2025 measurement \cite{ref28} applied a Nuclei-based JavaScript spider to ≈5 million applications and identified \textasciitilde42,000 exposed tokens across 334 secret types, providing the first industry-scale lower-bound on credentials in served JavaScript. The present paper is methodologically complementary: a smaller per-application benchmark with cross-vendor LLM-validated ground truth, designed to quantify the structural detection gaps a regex-driven sweep cannot reach.

\textbf{Secret scanning tool evaluation.} Basak et al.~\cite{ref16} conducted a comparative evaluation of nine secret detection tools (including GitHub Secret Scanner, Gitleaks, SpectralOps, and TruffleHog) against a benchmark dataset of real-world credential leaks, providing precision and recall measurements and identifying the primary sources of false positives and false negatives across tool categories. Their work establishes the limits of existing scanner coverage. This paper extends that analysis to a scanning layer those tools were not designed to address: the runtime-served application surface.

\textbf{Credential leakage in mobile and mini-app ecosystems.} Shi et al.~\cite{ref17} identified systematic credential exposure in mini-applications hosted within super-app platforms such as WeChat, finding 15 categories of vulnerable services in which credentials embedded in mini-app bundles enabled account hijacking and phishing. While the delivery mechanism differs from web SPA bundles, the structural cause is identical: build-time credential injection into client-delivered artifacts. The exploitation patterns and the tooling gap are the same.

\textbf{JavaScript bundle security.} Rack and Staicu~\cite{ref18} conducted an empirical study of JavaScript bundling practices across large-scale web deployments, analyzing bundle composition, dependency inclusion, and the security implications of bundler behavior. Their analysis of AST-level bundle reversibility is directly relevant to the detection methodology described in Section 8: understanding how bundled JavaScript can be reverse-engineered to recover credential context informs both attacker technique and defender detection strategy. Lauinger et al.~\cite{ref19} documented the widespread inclusion of outdated and vulnerable JavaScript libraries in web applications, establishing the broader pattern of client-side JavaScript as an understudied attack surface.

\textbf{Static analysis tool limitations.} Brito et al.~\cite{ref20} evaluated JavaScript static analysis tools against a dataset of 957 real-world vulnerabilities in npm packages, finding significant gaps in detection coverage across tool categories. Their findings on the limitations of AST-based static analysis for JavaScript are consistent with the shift-left scanning gaps documented in Section 6: static analyzers operating on pre-build source code cannot detect credentials that are introduced at build time or delivered at runtime.

\textbf{Client-side JavaScript as an attack surface.} Prior work on client-side JavaScript security has focused primarily on XSS vulnerabilities \cite{ref21}, third-party script inclusion risks, and subresource integrity enforcement. DAST tooling has evolved to cover injection vulnerabilities and authentication flaws in running applications. Neither category was designed to identify credentials embedded in served JavaScript bundles or injected into HTML at runtime. This paper maps that gap and presents a detection methodology suited to the runtime attack surface, addressing the structural limitations identified across the related work reviewed above.

\subsection{Terminology}\label{terminology}

\begin{itemize}
\tightlist
\item
  \textbf{Shift-left:} Security controls applied before or at the point of code commit, secret scanning, SAST, pre-commit hooks
\item
  \textbf{Shift-right:} Security controls applied after deployment, DAST, runtime monitoring, WAF
\item
  \textbf{SPA:} Single-Page Application. A web application where the entire frontend is delivered as a JavaScript bundle loaded once by the browser
\item
  \textbf{APIM:} Azure API Management. A gateway service that proxies and manages access to backend APIs
\item
  \textbf{BFF:} Backend for Frontend. A server-side proxy layer owned by the frontend team that holds service credentials and proxies requests to backend APIs
\end{itemize}

\subsection{Threat Model}\label{threat-model}

\textbf{Attacker profile and capabilities.} This paper models a network-capable attacker with no prior authentication to the target application, the same capabilities available to any web visitor, bug bounty researcher, or penetration tester. The attacker can load the application in a browser, inspect served HTTP responses including JavaScript bundles and JSON API replies, and issue authenticated API calls using any credentials recovered from those responses. The attacker does not need network interception capabilities, special tooling, or insider access. The attack surface is entirely passive: credentials are read from content the application actively delivers to every visitor.

\textbf{What is in scope.} Credentials embedded in client-delivered artifacts (JavaScript bundles, HTML source, statically served JSON configuration files, and JSON or XML API responses) that are accessible without prior authentication. Service credentials (Azure AD client secrets, APIM subscription keys) that are structurally reachable by this attacker model and that grant API access beyond what the current user session is authorized for. The exploitation chains in Sections 3 and 4 operate entirely within this model.

\textbf{What is out of scope.} Server-side credential stores (Azure Key Vault, environment variables not reflected in served content, database connection strings not returned in API responses). Attacks requiring network interception, active injection, or compromise of infrastructure components. Social engineering, phishing, and supply chain attacks. The structural gap documented in this paper is not a network-layer vulnerability. It is a consequence of credentials materializing in client-accessible artifacts through the deployment paths described in Section 5.

\begin{center}\rule{0.5\linewidth}{0.5pt}\end{center}

\section{Background, Azure AD, APIM, and Why These Credentials Matter}\label{background-azure-ad-apim-and-why-these-credentials-matter}

\subsection{The Azure AD + APIM Architecture}\label{the-azure-ad-apim-architecture}

Azure API Management acts as a reverse proxy gateway sitting in front of backend APIs. It handles authentication and authorization by validating tokens and subscription keys, rate limiting, request and response transformation, and analytics. The typical flow in an Azure-native SPA application is:

\begin{figure}[!htbp]
\centering
\pandocbounded{\includegraphics[keepaspectratio,alt={Azure AD + APIM request flow. The browser SPA authenticates via Azure AD using client credentials and calls backend APIs through the APIM gateway using both a Bearer token and a subscription key.},keepaspectratio,width=\textwidth,keepaspectratio]{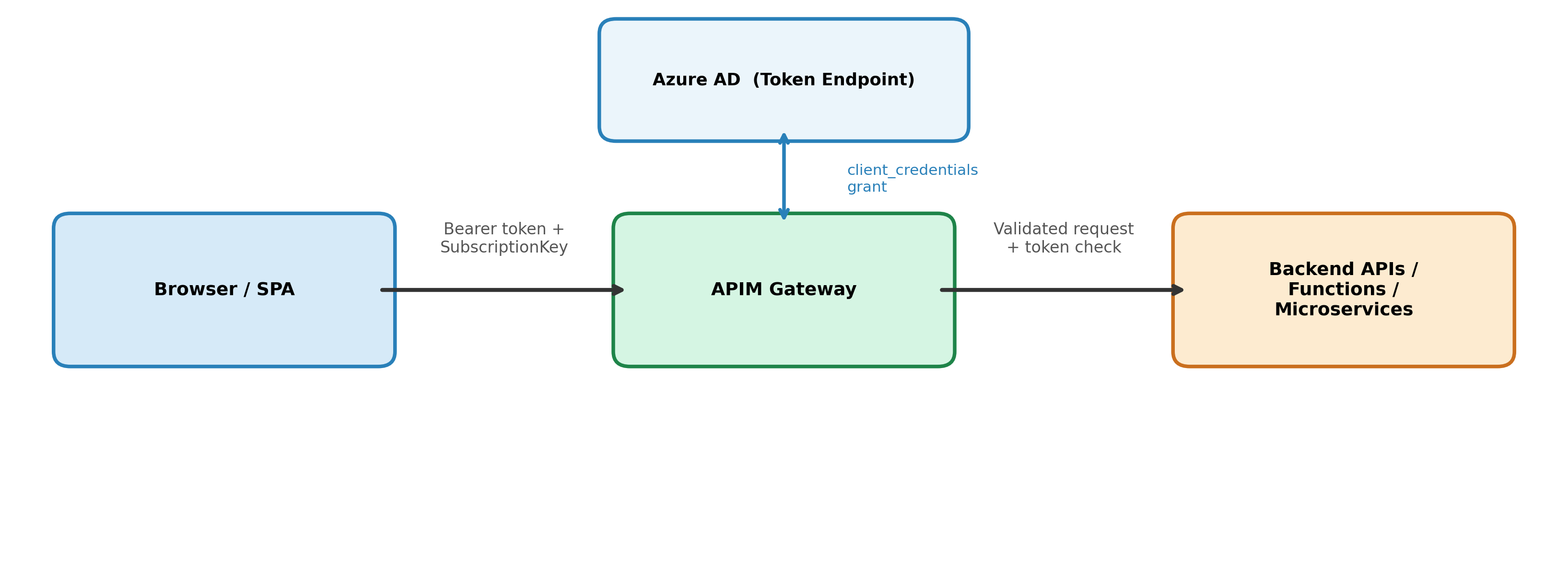}}
\caption{Azure AD + APIM request flow. The browser SPA authenticates via Azure AD using client credentials and calls backend APIs through the APIM gateway using both a Bearer token and a subscription key.}
\end{figure}

APIM enforces access through two separate credential types: Azure AD tokens (JWTs issued by the identity provider) and subscription keys (gateway-specific pass keys scoped to API products) \cite{ref7}. Both are required to reach protected endpoints.

\subsection{The Four Credential Values}\label{the-four-credential-values}

In the applications examined, four values were present in the client-side JavaScript bundle:

{\def\LTcaptype{none} 
\begin{table}[!ht]
\centering
\caption{The four Azure AD and APIM credential values present in the client-side bundle, their purpose, and whether each legitimately belongs in the browser.}\label{tab:credentials}
\adjustbox{max width=\textwidth}{%
\begin{tabular}{@{}
l
  l
  l@{}}
\toprule\noalign{}
Credential
 & Purpose
 & Should it be in the browser?
 \\
\midrule\noalign{}

\bottomrule\noalign{}

\textbf{AppID} (Client ID) & OAuth2 client identifier registered in Azure AD. Tells the identity provider which application is authenticating & Yes, public by OAuth2 spec for public clients \\
\textbf{AppKey} (Client Secret) & Password paired with AppID. Used in confidential client flows (server-to-server) & \textbf{Never} \\
\textbf{Resource} & API resource URI or scope used during token acquisition & Harmless alone, combined with AppKey enables full token generation \\
\textbf{SubscriptionKey} & APIM gateway pass key (\passthrough{\lstinline!Ocp-Apim-Subscription-Key!} header) required to reach APIM-protected APIs & \textbf{Never} \\
\end{tabular}%
}
\end{table}

}

The AppID is legitimately public in OAuth2 public client flows. The AppKey is not. It is the client secret for a confidential client flow, designed for server-to-server authentication where the application code is not visible to end users \cite{ref1, ref4}. Client secrets are architecturally valid for server-side web applications. The problem is not that the secret exists. It is that it appeared in browser-accessible JavaScript, where it is visible to every visitor who opens developer tools. A browser-based SPA is a public client and should not be issued a client secret at all; where it must authenticate users, the correct pattern is the OAuth 2.0 Authorization Code flow with Proof Key for Code Exchange (PKCE) \cite{ref29}, which requires no client secret (see Section~\ref{recommendations}). OAuth2 grant types operate independently on a single application registration. A client secret present on the registration activates the client credentials grant unconditionally, the authorization server applies no constraint between grant types, so the intended authorization flow does not restrict which grants the server will accept \cite{ref1}.

\subsection{Why All Four Values Together Is Critical}\label{why-all-four-values-together-is-critical}

The AppKey alone is insufficient: it requires the AppID and resource URI to generate a token. The SubscriptionKey alone is also insufficient, since API endpoints require a valid Bearer token. Individually, each value has limited reach. Together, they represent a complete authentication package: the ability to authenticate as the application itself and call any API the application is authorized to access.

The tenant ID (required to construct the token endpoint URL) was visible in each application's login redirect URL, hardcoded alongside the other values in the same bundle.

\begin{center}\rule{0.5\linewidth}{0.5pt}\end{center}

\section{Exploitation Chain 1, Azure AD + APIM}\label{exploitation-chain-1-azure-ad-apim}

\subsection{Discovery}\label{discovery}

The client-side JavaScript bundle of a production application (Azure AD authenticated, sitting behind an API Management gateway, serving thousands of users) contained all four credential values described in Section 2.2. The bundle was served over HTTPS to every visitor of the application with no authentication required to retrieve it.

The tenant ID was present in the application's login redirect URL, which was also hardcoded in the bundle.

\subsection{Token Generation}\label{token-generation}

The Azure AD token endpoint was called using the extracted credentials:

\begin{lstlisting}
POST https://login.microsoftonline.com/{tenant_id}/oauth2/token

grant_type=client_credentials
client_id={AppID}
client_secret={AppKey}
resource={Resource}
\end{lstlisting}

The endpoint returned a fully authenticated Bearer access token. The grant type used (\passthrough{\lstinline!client\_credentials!}) is a confidential client flow designed for server-to-server communication where the client secret is stored securely on the server \cite{ref1}. (The v1 endpoint format is shown, the v2 endpoint uses \passthrough{\lstinline!/oauth2/v2.0/token!} with a \passthrough{\lstinline!scope!} parameter in place of \passthrough{\lstinline!resource!}. The difference is that v1 identifies the target API by a single audience URI (\passthrough{\lstinline!resource!}), whereas v2 follows the OAuth 2.0 convention of naming granular per-permission scopes (for example \passthrough{\lstinline!api://<app-id>/.default!}); the underlying client-credentials grant is identical. Both endpoints remain in active use across enterprise Azure environments.) The application was using it in a public client context where the secret was visible to every user.

\subsection{API Endpoint Reconstruction}\label{api-endpoint-reconstruction}

Rather than scanning the bundle for additional secrets, it was read as a map of how the application was built. Buried in the minified code were API endpoint definitions, GET and POST endpoints with complete schema structures showing exactly which parameters each endpoint expected. The frontend had documented its own backend.

Using the Bearer token and APIM subscription key together:

\begin{lstlisting}
Authorization: Bearer {token}
Ocp-Apim-Subscription-Key: {SubscriptionKey}
\end{lstlisting}

The reconstructed endpoints were called. User profile data returned immediately. A password reset endpoint was identified using the same schema reconstruction approach, same token, same subscription key, same reconstructed schema. The finding was reported at this point and the assessment did not proceed further.

\subsection{The Scope Condition}\label{the-scope-condition}

The exposed credentials alone constitute a significant finding regardless of what they unlock. A client secret in a public JavaScript file means any visitor can authenticate as the application, the blast radius depends entirely on what permissions that application has been granted. In this case, the amplifying factor was that the service principal had been granted over-permissive API scopes. Application permissions (acting as the application itself, not as a specific user) had been granted where delegated permissions (acting as the signed-in user) would have been appropriate and sufficient \cite{ref3, ref4}.

The result: the application's client identity had been given the ability to perform user-level operations across every account in the system. Without this misconfiguration, the Bearer token would have had access only to application-level resources. With it, the token was effectively an administrative credential for every user account.

\subsection{Subscription Key Blast Radius}\label{subscription-key-blast-radius}

One APIM subscription key typically maps to a product containing multiple APIs:

\begin{lstlisting}
SubscriptionKey -> APIM Product "Internal APIs"
                     |-- /users/*
                     |-- /payments/*
                     |-- /admin/*
                     `-- /reports/*
\end{lstlisting}

One exposed subscription key grants access to every API in that product. The scope of exposure is determined by how the APIM product is configured. A detail that is not visible from the client side and must be assessed in the APIM portal.

\textbf{Result: Full account takeover from four values in a JavaScript file served to every visitor of the application.}

This was not a sophisticated attack. It required no exploit framework, no vulnerability scanner, and no special tooling. It required reading the JavaScript file the application was already serving to everyone and understanding what the credentials unlocked.

\begin{center}\rule{0.5\linewidth}{0.5pt}\end{center}

\section{Exploitation Chain 2, Client-Side Encryption Bypass}\label{exploitation-chain-2-client-side-encryption-bypass}

\subsection{Discovery}\label{discovery-1}

A different application, a different codebase, and a different obfuscation approach, but the same Azure AD credential pattern waiting at the end.

The credentials reached production via Path 1 (Section 5.1): Azure AD client credentials were stored in an environment file and baked into the JavaScript bundle at build time. Rather than leaving them as plaintext, the development team had encrypted the configuration object using CryptoJS before bundling. A pattern intended to obscure the credentials from casual inspection. The developer intent is that the configuration is protected because it is encrypted.

The problem: the decryption key was hardcoded in the same JavaScript file, three lines away from the encrypted string \cite{ref2, ref12}.

\subsection{Decryption}\label{decryption}

The hardcoded key was used to decrypt the environment string using the same CryptoJS call visible in the source:

\begin{lstlisting}
CryptoJS.AES.decrypt(encryptedConfig, hardcodedKey).toString(CryptoJS.enc.Utf8)
\end{lstlisting}

The output was a complete configuration object. Every Azure credential, every service key, every internal endpoint the application needed to function. The encryption had provided exactly zero protection. The key to unlock everything was sitting next to the lock.

The decrypted object contained the same pattern: AppID, AppKey, Resource, Azure AD client credentials embedded in what the development team believed was a secured configuration.

\subsection{Exploitation}\label{exploitation}

As with Chain 1, the service principal had been granted over-permissive scopes, Application permissions allowing an application-level token to perform user-level data access operations.

From the decrypted credentials:

\begin{itemize}
\tightlist
\item
  A Bearer token was generated using the extracted Azure AD client credentials
\item
  API endpoints were located hardcoded elsewhere in the bundle
\item
  Endpoint schemas were reconstructed from the minified bundle structure
\item
  Authenticated GET endpoints were called with the Bearer token
\item
  Full user profile data (names, email addresses, account identifiers) was returned for any user in the application
\end{itemize}

\textbf{Result: Personal data for thousands of users was accessible to anyone who opened the JavaScript file and understood what the encrypted string was hiding.}

\subsection{Why CryptoJS Obfuscation Is Particularly Dangerous}\label{why-cryptojs-obfuscation-is-particularly-dangerous}

The CryptoJS pattern creates a false sense of security that is difficult to identify in a standard code review. Developers implement it believing the configuration is protected. Security reviewers see encryption and move on. Static secret scanners see an encrypted string and find nothing to flag.

The construct is meaningless only when you read the entire file rather than scan it for plaintext secrets, which is exactly what static scanners do not do.

This pattern is not isolated to a single application or development team. Across the 113-application corpus assessed in this study, the CryptoJS obfuscation construct (encrypted configuration object with a co-located decryption key) recurred across multiple unrelated codebases, with 16 applications (14.2\% of vulnerable apps) containing CryptoJS-encrypted configurations and 37 distinct CryptoJS-AES encrypted-configuration blobs in the GT-194 benchmark. The recurrence across unrelated codebases suggests the pattern originates from a shared internal library, a shared development template, or a common architectural recommendation propagated across teams. A single flawed pattern adopted at the architecture or template level can introduce the same cryptographic misuse across an entire application portfolio simultaneously.

\begin{center}\rule{0.5\linewidth}{0.5pt}\end{center}

\section{How Secrets Reach Production}\label{how-secrets-reach-production}

Credentials reach production undetected through structural paths in modern build and deployment pipelines, regardless of organizational security maturity. Shift-left tools do not fail in these scenarios. The path from source code to production contains branches that no pre-deployment scanner covers. Five such paths are characterized in Sections 5.1 through 5.5.

\subsection{Path 1, Build-Time Environment Injection}\label{path-1-build-time-environment-injection}

React applications using \passthrough{\lstinline!REACT\_APP\_*!} environment variables, and Angular applications using \passthrough{\lstinline!environment.prod.ts!}, pass credentials to the build tool at compile time. webpack's DefinePlugin and the Angular CLI substitute these values directly into the output bundle \cite{ref6}.

The credential never exists in the repository. It is read from the build environment at compile time and written into the output artifact. Every secret scanner that ran before the build completed saw nothing, because there was nothing to see.

The Angular CLI's production build pattern deserves particular attention. The 113 credential-bearing applications break down by framework as follows:

{\def\LTcaptype{none} 
\begin{table}[!ht]
\centering
\caption{Front-end framework distribution across the assessed application corpus.}\label{tab:frameworks}
\adjustbox{max width=\columnwidth}{%
\begin{tabular}{@{}
lll@{}}
\toprule\noalign{}
Framework & Apps & Share \\
\midrule\noalign{}

\bottomrule\noalign{}

Angular (CLI) & 75 & 66\% \\
React & 25 & 22\% \\
REST API / JSON config & 5 & 4\% \\
Legacy / jQuery & 4 & 4\% \\
ASP.NET WebForms & 2 & 2\% \\
React (CRA) & 2 & 2\% \\
\end{tabular}%
}
\end{table}

}

Angular's dominance (66\%) is not coincidental. The Angular CLI treats \passthrough{\lstinline!environment.prod.ts!} as the documented and recommended approach for environment-specific configuration, credential injection is the framework's intended pattern. webpack's DefinePlugin structurally ensures that any value placed in the environment file is compiled directly into the output bundle. The prevalence of credential exposure across this portfolio is a direct consequence of framework architecture, not isolated developer error. Where a framework's recommended pattern produces credentials in client-delivered artifacts by design, the risk is systemic rather than individual.

\subsection{Path 2, CI/CD Pipeline Variable Substitution}\label{path-2-cicd-pipeline-variable-substitution}

This is a frequently observed path. A placeholder value lives in the repository while the real credential is stored as a pipeline variable in Azure DevOps, GitHub Actions, or a similar CI/CD platform. The placeholder is what every repository scanner, pre-commit hook, and SAST tool sees. The credential materializes only in the build artifact, after every scanner has already completed. It never exists in git at any point.

\begin{figure}[!htbp]
\centering
\pandocbounded{\includegraphics[keepaspectratio,alt={Two structural paths by which credentials reach production without repository exposure. Left, build-time environment injection: the credential is injected by webpack or the Angular CLI at compile time and never exists in git. Right, CI/CD pipeline variable substitution: a placeholder lives in source while the real credential is written into the artifact only after every scanner has completed.},keepaspectratio,width=\textwidth,keepaspectratio]{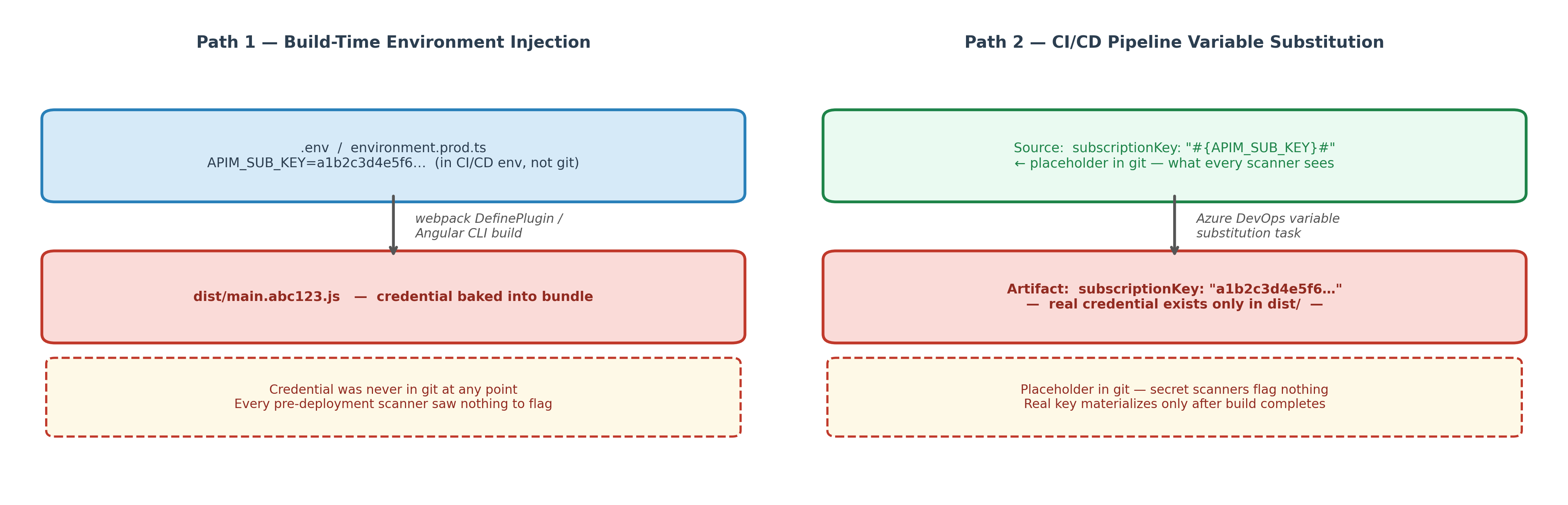}}
\caption{Two structural paths by which credentials reach production without repository exposure. Left, build-time environment injection: the credential is injected by webpack or the Angular CLI at compile time and never exists in git. Right, CI/CD pipeline variable substitution: a placeholder lives in source while the real credential is written into the artifact only after every scanner has completed.}
\end{figure}

\subsection{Path 3, Runtime Configuration Injection}\label{path-3-runtime-configuration-injection}

Some applications fetch their configuration after the browser loads the initial bundle:

\begin{itemize}
\tightlist
\item
  A \passthrough{\lstinline!/assets/config.json!} file served statically at runtime alongside the application
\item
  A \passthrough{\lstinline!window.\_\_APP\_CONFIG\_\_!} object injected into \passthrough{\lstinline!index.html!} by the server at request time
\item
  SSR state blobs (the \passthrough{\lstinline!\_\_NEXT\_DATA\_\_!} object injected by Next.js or \passthrough{\lstinline!window.\_\_INITIAL\_STATE\_\_!} injected by Nuxt) which regularly contain tokens, API endpoints, and internal service configuration
\end{itemize}

These configurations are loaded after deployment, sometimes from CDN nodes, sometimes from origin servers. They do not exist as files in the repository. They are assembled and served at runtime, invisible to any scanner that ran before deployment.

In the applications assessed for this study, two targets were REST API configuration endpoints (not JavaScript bundles) returning credential-bearing JSON responses directly to the browser. In both cases the endpoint was called by the SPA on load to retrieve runtime configuration, and the response contained Azure AD client credentials in plaintext JSON fields. These endpoints were publicly accessible without authentication. This confirms that the runtime attack surface extends beyond JavaScript files to include configuration APIs that serve secrets to the browser on demand, and that no JavaScript-only scanner would detect credentials delivered through this path.

\subsection{This Pattern Extends Beyond Azure}\label{this-pattern-extends-beyond-azure}

The exploitation chains documented in this paper involve Azure AD and Azure API Management because that is the stack present in the applications assessed. The same structural paths appear across other providers. The remediation guidance in Section 9 is Azure-specific and should be adapted to the relevant platform.

The same paths appear across every stack:

\begin{itemize}
\tightlist
\item
  \textbf{AWS:} Cognito user pool client IDs and client secrets embedded in React bundles via \passthrough{\lstinline!REACT\_APP\_*!} variables. API Gateway API keys substituted into JavaScript at build time. S3 presigned URL generation credentials hardcoded in SPA configuration.
\item
  \textbf{GCP:} Firebase API keys and project configuration objects (\passthrough{\lstinline!apiKey!}, \passthrough{\lstinline!authDomain!}, \passthrough{\lstinline!projectId!}) served in \passthrough{\lstinline!firebase-init.js!} to every visitor. These are legitimately public for Firebase Authentication but are frequently accompanied by service account credentials that are not.
\item
  \textbf{Generic SaaS:} Twilio account SIDs and auth tokens, SendGrid API keys, Stripe secret keys, and Mailgun API keys substituted into build artifacts from CI/CD pipeline variables. The credential format changes. The path to production does not.
\end{itemize}

The exploitation technique is identical across all of these: read what the application serves, identify the credential, authenticate with it. The shift-left tooling gap is identical: none of these credentials touched a repository at any point in their path to the browser.

\subsection{Credentials Persist Across All Environments}\label{credentials-persist-across-all-environments}

Because credential injection occurs at build time, every environment that runs a build receives the same secrets baked into its artifact. Development, SIT, UAT, pre-production, and production environments all receive credentials through the same pipeline, and all produce the same exposure.

Across the applications assessed in this study, the same credential set was confirmed present in production, UAT, SIT, and pre-production environments of the same application in multiple cases. Lower environments typically have weaker access controls, broader team access, and no security monitoring, making them at least as attractive a target as production for an attacker who has identified the pattern.

This has a direct consequence for remediation: rotating credentials in production without simultaneously fixing the build pipeline leaves every lower environment still exposed. An attacker who has already extracted credentials from a UAT environment retains access regardless of what happens in production. Remediation is only complete when the credential is removed from the build pipeline and rotated across all environments simultaneously.

\subsection{Path 4, Third-Party Script Inclusion (Supply-Chain Credentials)}\label{path-4-third-party-script-inclusion-supply-chain-credentials}

Modern web applications routinely embed third-party scripts. Analytics platforms, tag managers, customer-support widgets, error monitoring SDKs, advertising libraries, and CDN-hosted dependencies all execute in the application's origin and run with full access to its DOM, cookies, and storage. These scripts also carry their own credentials, including API keys, project identifiers, instrumentation tokens, and customer-tenant identifiers that are public-by-design from the third party's perspective but materialize in the served bundle of every site that integrates them.

Two structural problems follow. First, the credentials embedded by third parties are outside the application owner's repository, so no scanner the owner runs can flag them. They enter the browser through a script tag that the owner did not author. Second, the security posture of those third-party endpoints, including rate limiting, scope enforcement, and audit logging, is determined by the third party rather than the application owner. Demir et al.~\cite{ref15} reported that 16\% of verified credential exposures originated from third-party inclusions, with the highest third-party rates observed for OpenAI (24\%), Twilio (24\%), Mailchimp (23\%), Alibaba (23\%), Stripe (22\%), and SendGrid (22\%) credentials served via embedded SDKs that the integrating application did not control. In the corpus assessed for this study, the Google API keys (6 instances in GT-194) and a subset of the Other API key/token class are consistent with this path, integrated through third-party SDKs whose credential material enters the served bundle outside any first-party scanning workflow. The exposure pattern is structurally indistinguishable from build-time or runtime injection. A credential reaches a deployed artifact, is served to every visitor, and has no automated detection on either side of the integration.

\subsection{Path 5, Scanner Suppression at the Organizational Level}\label{path-5-scanner-suppression-at-the-organizational-level}

A fifth contributing factor operates at the organizational level and compounds the four structural paths above. Many SPA architectures structurally require credentials in the browser, including Google Maps API keys, Firebase configuration, Stripe publishable keys, and Twilio client tokens. When shift-left scanners flag these, the operational path of least resistance is to suppress the alert through ignore-listing files, dismissing alerts as false positives, adding inline suppression comments, or allowlisting specific patterns in the SAST ruleset. The suppression is deliberate. The developer knows the credential is real but believes it is an accepted architectural risk or lacks the authority or time to implement the server-side proxy pattern (Section 9.1) that would eliminate the in-browser requirement entirely. The credential is live, required, and invisible to the security tooling stack. The build process did not bypass the scanner. The scanner was deliberately told to look away. The outcome is identical to the structural paths. A live credential reaches a production artifact with no automated monitoring.

\begin{center}\rule{0.5\linewidth}{0.5pt}\end{center}

\section{Why Shift-Left Scanners Miss This}\label{why-shift-left-scanners-miss-this}

The five paths in Section 5 produce credentials in deployed artifacts through different mechanisms but converge at the same boundary. Shift-left scanners run before or at the point of commit, before the build executes, before the pipeline substitutes variables, before any third-party script is fetched, and before the artifact is assembled. The credential does not exist when the scanner runs. By the time it exists in the served artifact, every shift-left scanner has already completed.

The table below maps this timing mismatch precisely.

{\def\LTcaptype{none} 
\begin{table}[!ht]
\centering
\caption{Where each credential is present across the delivery pipeline, and which scanning layer, shift-left or shift-right, covers each stage.}\label{tab:pipeline-gap}
\adjustbox{max width=\columnwidth}{%
\begin{tabular}{@{}
l
  l
  l
  l@{}}
\toprule\noalign{}
Pipeline Stage
 & Credential present?
 & Shift-left scans here?
 & Shift-right covers here?
 \\
\midrule\noalign{}

\bottomrule\noalign{}

Source code / git repository & No, placeholder or absent & Yes & No \\
CI/CD pipeline variable store & No, stored server-side & No & No \\
Build artifact (\passthrough{\lstinline!dist/!}) & \textbf{Yes, injected at compile time} & No & No \\
Third-party SDK delivered at runtime & \textbf{Yes, embedded by integration} & No & No \\
Deployed to CDN / Blob Storage & \textbf{Yes, served to every visitor} & No & No \\
Runtime-fetched config (\passthrough{\lstinline!/assets/config.json!}) & \textbf{Yes, fetched by browser} & No & No \\
\end{tabular}%
}
\end{table}

}

The gap is not merely a configuration failure or a tool weakness. It is a structural property of when each scanning layer operates relative to when the credential materializes. More aggressive rule tuning may narrow the static-tool gap on credentials that are present in the artifact (§10), but no amount of pre-deployment rule tuning can reach a credential that only materializes at runtime; closing that portion of the gap requires the scanning layer itself to change.

\begin{figure}[!htbp]
\centering
\pandocbounded{\includegraphics[keepaspectratio,alt={Where shift-left scanning stops versus where secrets materialize. Shift-left tooling covers only the source code layer. Secrets introduced through build-time injection, pipeline variable substitution, runtime configuration fetching, or third-party script inclusion exist only in deployed artifacts and served content. These are layers no pre-deployment scanner reaches.},keepaspectratio,width=\textwidth,keepaspectratio]{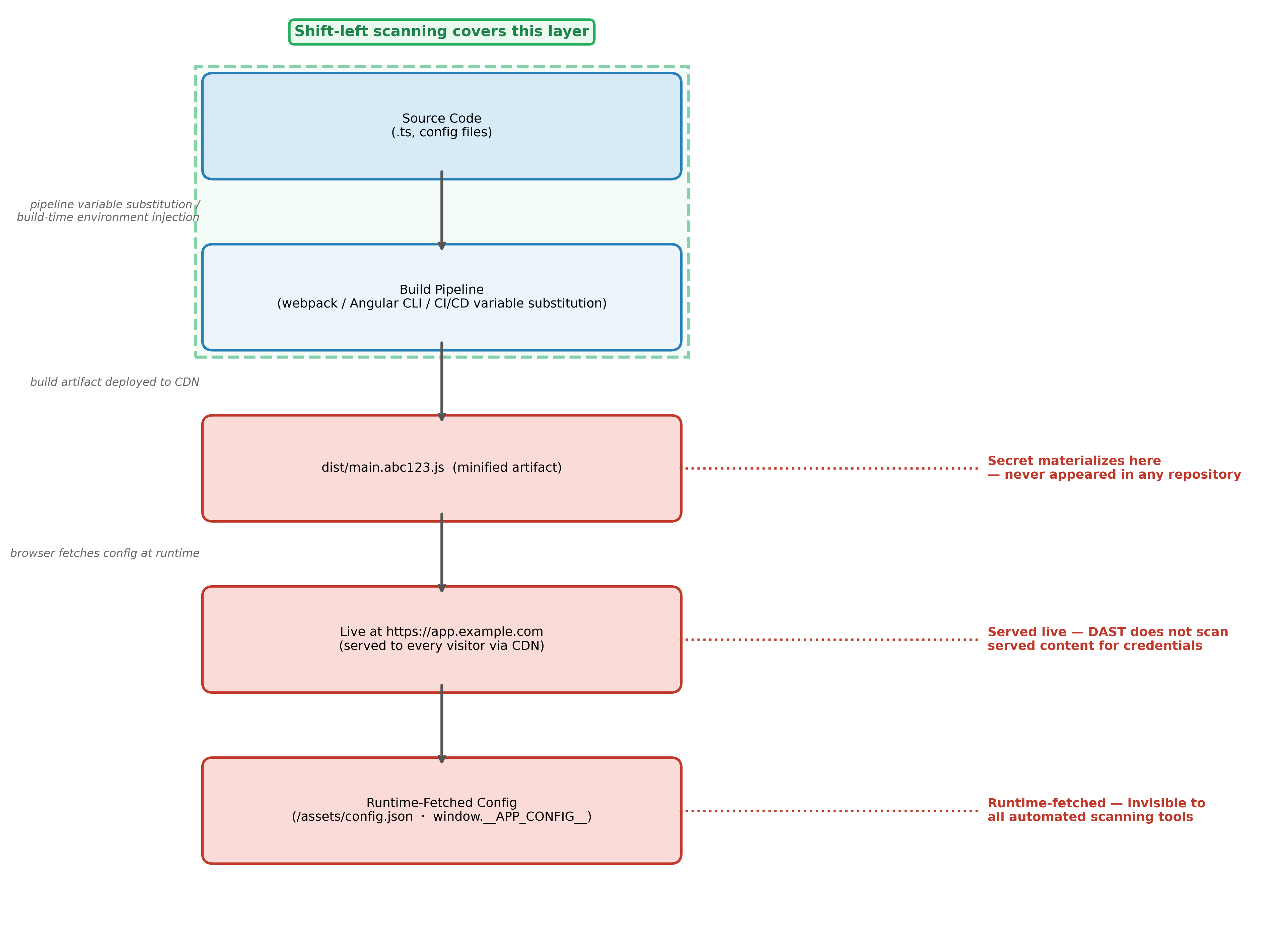}}
\caption{Where shift-left scanning stops versus where secrets materialize. Shift-left tooling covers only the source code layer. Secrets introduced through build-time injection, pipeline variable substitution, runtime configuration fetching, or third-party script inclusion exist only in deployed artifacts and served content. These are layers no pre-deployment scanner reaches.}
\end{figure}

The shift-left tooling market is mature. GitLeaks, TruffleHog, detect-secrets, and GitHub Advanced Security are widely deployed \cite{ref8}. SAST tools (Semgrep, SonarQube, Checkmarx) are standard in enterprise CI/CD pipelines. Each tool does exactly what it is designed to do. None of them are designed to scan build artifacts, third-party SDK output, or live deployed applications.

{\def\LTcaptype{none} 
\begin{table}[!ht]
\centering
\caption{Representative shift-left scanners, what each scans, and why each misses credentials that materialize after the source-code stage.}\label{tab:shiftleft}
\adjustbox{max width=\textwidth}{%
\begin{tabular}{@{}
l
  l
  l@{}}
\toprule\noalign{}
Scanner
 & What It Scans
 & Why It Misses
 \\
\midrule\noalign{}

\bottomrule\noalign{}

\textbf{GitLeaks} & Git commits, staged files, history & Pipeline variables are substituted after git. Placeholder is in git, real value is not. Never sees \passthrough{\lstinline!dist/!} \\
\textbf{SAST (Checkmarx, SonarQube, Veracode)} & Source \passthrough{\lstinline!.ts/.js!} pre-build & Sees \passthrough{\lstinline!process.env.SUBSCRIPTION\_KEY!}, no secret present. Does not execute the build \\
\textbf{Semgrep / CodeQL} & AST of source code & Pre-build source only, same gap as SAST \\
\textbf{GitHub Advanced Security} & Repository content & \passthrough{\lstinline!$\{\{ secrets.KEY \}\}!} injected into build output never touches the repository \\
\end{tabular}%
}
\end{table}

}

\subsection{The Minification Problem}\label{the-minification-problem}

Pattern-based secret scanners rely on variable names and known key formats to identify secrets. Minification destroys variable names:

\begin{lstlisting}
// Source, scanner identifies this:
const subscriptionKey = "abc123xyz789..."

// Minified, scanner may miss this:
const a={b:"abc123xyz789..."}
\end{lstlisting}

Entropy-based detection addresses part of this problem (high-entropy strings are flagged regardless of variable name) but entropy alone generates significant noise against the minified JavaScript of a modern SPA, which contains many high-entropy strings that are not credentials (hashed asset filenames, base64-encoded resources, obfuscated library code).

\subsection{Source Maps, The Silent Bypass}\label{source-maps-the-silent-bypass}

Angular and webpack generate \passthrough{\lstinline!.map!} files alongside minified bundles by default:

\begin{lstlisting}
dist/main.abc123.js        <- minified
dist/main.abc123.js.map    <- full original source, variable names, logic intact
\end{lstlisting}

If source maps are deployed to production (a common default) any credential ``hidden'' in minified JavaScript is fully readable in browser developer tools via the Sources panel. SAST scans the \passthrough{\lstinline!.ts!} source. The \passthrough{\lstinline!.map!} file exposes everything at runtime regardless of what the scanner found. Disabling source maps in production builds (\passthrough{\lstinline!sourceMap: false!} in \passthrough{\lstinline!angular.json!}) removes this bypass entirely.

\begin{center}\rule{0.5\linewidth}{0.5pt}\end{center}

\section{Why Shift-Right Scanners Also Miss This}\label{why-shift-right-scanners-also-miss-this}

Dynamic Application Security Testing tools receive JavaScript files as part of application spidering, but they treat those files as attack delivery vehicles, not as targets. DAST tools spider the application looking for injection points, authentication flaws, and misconfigurations. They do not parse \passthrough{\lstinline!main.js!} for \passthrough{\lstinline!Ocp-Apim-Subscription-Key!} patterns.

The shift-right tooling landscape by category:

\begin{figure}[!htbp]
\centering
\pandocbounded{\includegraphics[keepaspectratio,alt={Approximate tooling coverage by runtime security category. Mature tooling exists for injection vulnerabilities and network anomalies. Deployed secrets in static assets and served content represent a largely unsolved coverage gap.},keepaspectratio,width=\textwidth,keepaspectratio]{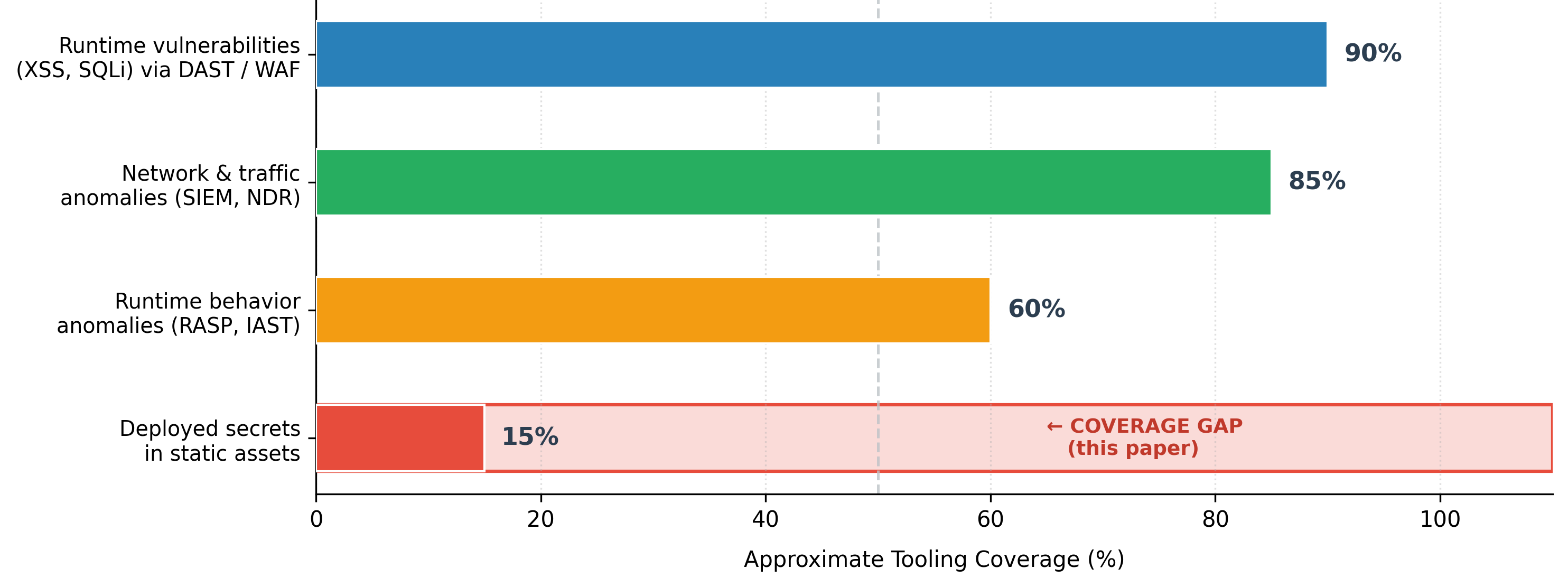}}
\caption{Approximate tooling coverage by runtime security category. Mature tooling exists for injection vulnerabilities and network anomalies. Deployed secrets in static assets and served content represent a largely unsolved coverage gap.}
\end{figure}

Tools that partially address this class of finding:

{\def\LTcaptype{none} 
\begin{table}[!ht]
\centering
\caption{Representative shift-right tools that partially address deployed-secret detection, their approach, and their limitations.}\label{tab:shiftright}
\adjustbox{max width=\textwidth}{%
\begin{tabular}{@{}
l
  l
  l@{}}
\toprule\noalign{}
Tool
 & Approach
 & Limitation
 \\
\midrule\noalign{}

\bottomrule\noalign{}

\textbf{Nuclei} & Community templates matching secret patterns in HTTP responses & Regex-based, misses minified variable names, template coverage varies \\
\textbf{TruffleHog} (filesystem/web mode) & Entropy and pattern detection on files & Web mode experimental. Reliable on artifact filesystem, not live application scanning \\
\textbf{Detectify / Probely} & DAST with secret detection, downloads JS assets & Does not execute JavaScript, misses runtime-fetched configurations \\
\textbf{GitGuardian perimeter} & Monitors public-facing endpoints & Focused on public GitHub exposure, not internal deployed applications \\
\end{tabular}%
}
\end{table}

}

None of these tools provide comprehensive coverage of the full runtime attack surface of a modern SPA: split webpack bundle chunks loaded dynamically, SSR state objects injected into HTML, JSON and XML API responses returning credentials in fields never meant to be client-facing, and outbound request headers carrying subscription keys and bearer tokens on every API call.

\textbf{The result: once a secret evades shift-left controls and reaches production, it is effectively invisible to automated tooling. The only things finding runtime secrets in production today are manual penetration testers, bug bounty researchers, and attackers. Two of those three report what they find.}

\begin{center}\rule{0.5\linewidth}{0.5pt}\end{center}

\section{What Runtime Layer Detection Requires}\label{what-runtime-layer-detection-requires}

Closing the shift-right gap requires a different scanning model than the tools currently deployed. This section describes what a tool operating at the runtime layer must address, the attack surface it must cover, the detection approach required for minified production JavaScript, and the operational constraints that determine whether findings are actionable in practice.

\subsection{Attack Surface Coverage}\label{attack-surface-coverage}

A runtime-layer scanner must cover what the application actually serves, not what exists in its source repository. The full attack surface of a modern web application includes:

\begin{itemize}
\tightlist
\item
  \textbf{JavaScript bundles}, including split webpack chunks that modern SPAs load dynamically as users navigate, not just the initial bundle
\item
  \textbf{HTML source}, SSR state blobs (\passthrough{\lstinline!\_\_NEXT\_DATA\_\_!}, \passthrough{\lstinline!window.\_\_INITIAL\_STATE\_\_!}) injected by server-side rendering frameworks, which regularly contain tokens and internal configuration \cite{ref3}
\item
  \textbf{JSON API responses}, credentials returned in fields never intended to be client-facing, which static scanners never see
\item
  \textbf{XML responses}, enterprise services carrying connection strings and service credentials in structured response bodies
\item
  \textbf{Outbound request headers}, subscription keys and bearer tokens sent on every API call in plaintext
\end{itemize}

For bulk assessment, following \passthrough{\lstinline!<script src>!} references and tracing chunk URLs is required to scan an application's full JavaScript surface rather than only the files present in the initial page load.

\subsection{Detection Approach}\label{detection-approach}

Production JavaScript presents a different detection problem than source code. A layered approach is required:

\textbf{Anchored vendor token patterns}, Known credential formats from cloud providers (Azure, AWS, GCP), SaaS platforms (Twilio, SendGrid, Stripe), and authentication services have identifiable structure. Anchored regex patterns matched against specific token formats produce fewer false positives than generic high-entropy matching and are resistant to minification because they target the value format, not the variable name.

\textbf{Shannon entropy analysis} (For credentials without known formats) internal API keys, session tokens, symmetric encryption keys, Shannon entropy identifies high-randomness strings that are statistically unlikely to be application logic \cite{ref11}. Entropy analysis catches credentials that minification has stripped of their variable name context. However, entropy analysis alone against minified JavaScript generates significant noise: hashed asset filenames, base64-encoded resources, and obfuscated library code all produce high-entropy strings that are not credentials.

\textbf{Key-value context scanning}, Examining the structural context surrounding a high-entropy value reduces noise by requiring that the value appear in a credential-relevant context. A JSON key named \passthrough{\lstinline!subscriptionKey!}, \passthrough{\lstinline!apiKey!}, or \passthrough{\lstinline!clientSecret!} paired with a high-entropy string is significantly more likely to be a credential than an isolated high-entropy string. This context requirement is the primary mechanism for reducing false positives in minified JavaScript where variable names are not available.

\textbf{PII detection with validation}, Pattern matching alone is insufficient for PII. Credit card numbers require Luhn algorithm validation \cite{ref14}. Social Security Numbers require format and checksum verification. Without validation, the false positive rate for these patterns in arbitrary JavaScript is too high to be actionable.

\subsection{Operational Constraints}\label{operational-constraints}

Two operational constraints determine whether runtime scanning findings are actionable in practice:

\textbf{No outbound verification calls.} Verifying whether an exposed API key is valid by calling the issuing service generates log entries at the target provider, can trigger security alerts, and reveals the assessment to defenders. Runtime scanning must run locally with no outbound verification calls to third-party APIs or services.

\textbf{Noise suppression is a first-class requirement.} A scanner that generates hundreds of false positives per scan is a scanner that gets disabled. In the practitioner workflows where runtime scanning is most needed (penetration testing engagements, bug bounty assessments, security reviews) findings must be trustworthy enough to act on immediately. Every false positive erodes that trust and increases the likelihood that a real credential is dismissed as noise. Noise suppression is not a secondary quality concern. It is what determines whether the tool gets used. The detection methodology described in this section informed the design of SecretSifter, referenced in Section 9.3.

\textbf{Algorithm 1: Runtime Credential Detection}

\begin{figure}[!htbp]
\noindent\rule{\columnwidth}{0.5pt}\\
\textbf{Algorithm 1: Runtime Credential Detection}\\
\rule{\columnwidth}{0.3pt}
\begin{lstlisting}
Input:  URL , target application URL
Output: F   , set of (credential_type, value, context) findings

1.  F <- {}
2.  R <- HTTP_GET(URL, follow_redirects=true)
3.  C <- extract_content_units(R)
        // C = {JS bundles, HTML blobs, JSON responses, headers}
4.  for each chunk_url in extract_script_srcs(R.html) do
5.      C <- C U {HTTP_GET(chunk_url)}
6.  end for
7.
8.  for each content_unit u in C do
9.      // Layer 1: Anchored vendor token patterns
10.     for each pattern p in VENDOR_PATTERNS do
11.         matches <- regex_findall(p.regex, u)
12.         for each m in matches do
13.             F <- F U {(p.credential_type, m, surrounding_context(m, u))}
14.         end for
15.     end for
16.
17.     // Layer 2: Shannon entropy + key-value context
18.     tokens <- tokenize_kv_pairs(u)
19.     for each (key, value) in tokens do
20.         if shannon_entropy(value) >= ENTROPY_THRESHOLD then
21.             if key in CREDENTIAL_KEY_NAMES then
22.                 F <- F U {(infer_type(key), value, key)}
23.             end if
24.         end if
25.     end for
26.
27.     // Layer 3: PII with validation
28.     for each pattern p in PII_PATTERNS do
29.         matches <- regex_findall(p.regex, u)
30.         for each m in matches do
31.             if p.validator(m) = true then
32.                 F <- F U {(p.credential_type, m, surrounding_context(m, u))}
33.             end if
34.         end for
35.     end for
36. end for
37.
38. return deduplicate(F)
\end{lstlisting}
\noindent\rule{\columnwidth}{0.5pt}
\end{figure}

\emph{VENDOR\_PATTERNS includes anchored regex for Azure AD (client\_id, client\_secret), APIM subscription keys, AWS access keys, Google API keys, and 40+ additional provider formats. ENTROPY\_THRESHOLD = 3.5 bits/char. CREDENTIAL\_KEY\_NAMES includes apiKey, subscriptionKey, clientSecret, appKey, encryptionKey, and synonyms. No outbound verification calls are made at any step.}

\subsection{Preliminary Case Study: Detection Outcomes (N=2)}\label{preliminary-case-study-detection-outcomes-n2}

This section reports detection outcomes from the two confirmed-positive applications in the authorized assessments described in Section 1. This is a preliminary case study with N=2 confirmed-positive applications, not a controlled benchmark evaluation. No ground-truth labeled corpus of production JavaScript exists for this class of finding. Constructing one from unauthorized applications is not ethically feasible. What follows is a factual account of what the runtime detection methodology identified in both confirmed-positive cases, and why no other automated control was positioned to find the same.

\textbf{Credentials in the detection gap.} In both confirmed-positive applications, the exposed credentials had reached production through the structural paths described in Section 5, build-time environment injection and CI/CD pipeline variable substitution. By the time the credentials were present in the served application, they had passed every automated control in the pre-deployment pipeline. This is not a statement about whether those controls ran or how well they were configured. It is a structural property of the paths: credentials introduced after the build scanner runs, or substituted by the pipeline after the repository scanner completes, do not exist in any layer those tools examine. The shift-left scanning layer had nothing to find because the credentials were not there when it looked.

\textbf{No shift-right control covered this layer.} Once deployed, both applications were running in production environments with standard enterprise security controls, including web application firewalls and network monitoring. None of those controls are designed to examine the content of served JavaScript files for embedded credentials. DAST tools spider applications for injection vulnerabilities, WAF and network monitoring tools inspect inbound traffic for attack patterns. Neither category asks whether the JavaScript being served to users contains an Azure AD client secret. The credentials were live, publicly served, and invisible to the entire deployed security stack, not because any tool failed, but because no tool in any category was scanning that layer.

\textbf{Runtime detection results.} The layered detection approach described in Sections 8.1--8.2 (anchored vendor token pattern matching combined with key-value context scanning) identified the complete credential sets in both confirmed-positive applications. In both cases, the credentials were located in served JavaScript bundles. In Chain 1 (Section 3), the AppID, AppKey, Resource, and SubscriptionKey were matched directly by anchored Azure AD and APIM credential patterns. In Chain 2 (Section 4), the encrypted configuration was located by key-value context scanning identifying a high-entropy value paired with a credential-relevant key name. Decryption was performed manually using the co-located key, after which the same credential patterns applied. Both findings were confirmed by successful exploitation before disclosure.

\textbf{Credential set completeness.} Of the 113 applications with confirmed credential exposure, 63 (55.8\%) contained a complete Azure AD credential set (client\_id, client\_secret, tenant\_id, and resource URI co-located in the same JS bundle) sufficient to execute the exploitation chain documented in Section 3 given over-permissive service principal scopes. The remaining 50 applications contained partial credentials. Across all 113 vulnerable applications the most common credential types observed were Azure App Insights instrumentation keys (62 apps, 54.9\%), Azure APIM subscription keys (46 apps, 40.7\%), JWT tokens issued to the browser (25 apps, 22.1\%), CryptoJS-encrypted configuration objects (16 apps, 14.2\%), and CyberArk AIM tokens (11 apps, 9.7\%). Applications frequently contained multiple credential types simultaneously. Partial credentials represent significant exposure but do not independently enable the full account takeover chain. The high proportion of complete Azure AD sets (56\% of vulnerable applications) indicates that the credential injection pattern typically carries the full credential bundle rather than isolated values. Per-app prevalence and per-credential GT-194 counts can diverge in either direction (for example, 62 apps share 15 unique App Insights workspace iKeys, while the 16 CryptoJS-using apps contain 37 distinct encrypted blobs in GT-194).

\textbf{Scope and limitations.} The sample of 113 applications is drawn from a single authorized engagement scope of approximately 2,000 enterprise web application assets, not a randomly selected population across multiple organizations. Detection was performed by a skilled practitioner applying the methodology described in this section, not by a fully automated tool running without guidance. A formal evaluation (with a labeled benchmark corpus, automated tool execution, and precision and recall measurements at scale) remains an area for future empirical work. The outcomes reported here establish that the methodology successfully identifies credentials in the layer that existing tooling cannot reach, and that this layer contains real, exploitable credentials in production systems that have passed all deployed automated controls.

\subsection{Tool Comparison Study, GT-194 Cross-Vendor LLM-Validated Benchmark}\label{tool-comparison-study-gt-194-cross-vendor-llm-validated-benchmark}

Section 8.4 established that the runtime detection methodology successfully identifies credentials in two confirmed-positive applications. This section quantifies how the nine production scanners (established static tools at default and tightened-custom configurations, and the runtime-aware extension SecretSifter) perform against an independent Ground Truth, GT-194, on the same corpus, with the LLM-assisted ground-truth reference comparator reported separately. The evaluation is deliberately structured to remove the developer-author conflict of interest from the headline numbers: the Ground Truth is constructed by an LLM with no access to any evaluated detector, the LLM-extracted candidates were independently classified by an LLM from a different vendor, and the field-deployment register was locked on 8 May 2026.

\subsubsection{Benchmark Construction}\label{benchmark-construction}

\textbf{What ``GT-194'' means.} GT-194 is the Ground Truth set of 194 unique secret-grade credentials assembled from production JavaScript bundles across 113 enterprise applications. ``Ground Truth'' here means the labelled, manually validated set of credentials that scanners are measured against, the canonical denominator for recall calculations throughout this section. GT-194 includes only credential instances that are actually secret-bearing, partitioned into eight credential types: Azure APIM subscription key (n=54), Azure AD client\_secret incl.~btoa-encoded OAuth (n=50), CryptoJS-AES Salted\_\_ blob (n=37), Other API key/token incl.~App Insights iKey (n=29), JWT (n=11), Google API key (n=6), CyberArk AIM (n=4), and plaintext user credential (n=3). The eight type counts sum to 194.

\textbf{Corpus.} The 113-application benchmark is the full set of credential-bearing applications identified during the engagement (drawn from approximately 2,000 enterprise web application assets in scope). It includes all applications providing downloadable JavaScript bundles suitable for both static-scanner evaluation and LLM-assisted ground-truth extraction, and captures the full credential-bearing production deployment set including builds containing CryptoJS-AES blobs, Other API keys, and plaintext user credentials.

\textbf{Independent Ground Truth extraction.} Claude Opus 4.7 (Anthropic) \cite{ref22} was prompted to extract every value from the bundles that it considered a real high-impact secret. The model was given only the bundle content and a strict-secret rubric, with no access to SecretSifter rules, output, or any other tool's findings. After human review, the locked GT-194 was assembled as the union of this independent extraction and a manual analyst pass that added secret-grade credentials the model did not surface. Because the ground truth therefore contains credentials no single detector produced on its own, no scanner reaches full recall against it: even the extracting reference (Claude Opus) recovers only 166 of 194 (85.6\%, Table~\ref{tab:recall}), and the 27 credentials in the tool-agnostic blind spot (\S8.5.11) are those surfaced solely by manual analysis, detected by none of the nine evaluated production scanners and missed by the GT-construction reference comparator. Each of these 27 manual-only credentials was verified as a secret-grade credential by analyst review against the same strict-secret rubric; because they were not produced by the LLM extractor, the automated cross-vendor (GPT-5.5) validation reported below applies to the LLM-extracted candidate set, and the 27 rest on manual verification rather than on the automated validator. The blind-spot arithmetic reconciles as follows: the reference comparator misses 28 of 194 (194 minus 166); of those, one is recovered by SecretSifter (the single credential SecretSifter catches that the reference misses), leaving 27 that no production scanner recovers, which is the tool-agnostic blind spot. Finally, GT-194 covers first-party application JavaScript bundles only; it does not evaluate scanner performance against Path 4 (third-party script inclusion, \S5) credentials, whose remediation ownership and detection surface differ.

\textbf{Cross-vendor independent validation.} Each Opus-extracted candidate was independently classified by GPT-5.5 (OpenAI) \cite{ref23} via the OpenAI Codex CLI. GPT-5.5 received the candidate value, key name, and source-file context, and was asked to apply the same strict-secret rubric independently. Across the 247 LLM-extracted candidates, GPT-5.5 returned 207 concordant SECRET classifications, giving Brennan-Prediger \cite{ref24} κ = 0.676, substantial on the Landis-Koch interpretive scale \cite{ref26}. The 27 manual-only additions to GT-194 were not produced by the LLM extractor and are excluded from the κ denominator; they rest on analyst review against the same strict-secret rubric. Because the adjudicated candidate set is single-class in the SECRET label (every candidate is an Opus-labelled SECRET), Cohen's κ is undefined here (the Feinstein-Cicchetti paradox \cite{ref25}); Brennan-Prediger κ is the bias-corrected metric appropriate for single-class prevalence.

\textbf{Why public-by-design Azure identifiers are tracked separately (``below the line'').} The corpus also contains 249 chain-completion identifiers, 153 Azure AD App\_IDs / client\_ids / tenant\_ids and 96 Azure AD resource URIs (\passthrough{\lstinline!resource!}, \passthrough{\lstinline!Resource!}, \passthrough{\lstinline!*\_Resource!}, \passthrough{\lstinline!resourceId!}). These are public-by-design Azure identifiers, not credentials, and are therefore excluded from the GT-194 recall denominator. Two reasons drive the structural choice. First, including them would artificially inflate scanner scores, since these are not credentials a peer reviewer would accept as ``leaks.'' Second, excluding them entirely would hide the operational reality that all four components (\passthrough{\lstinline!client\_secret!}, \passthrough{\lstinline!client\_id!} / App\_ID, \passthrough{\lstinline!tenant\_id!}, and \passthrough{\lstinline!resource!} URI) must be combined to mint an Azure AD access token. Listing them separately preserves both honesty about the recall metric and visibility into the exploitation chain. The chain-completion identifiers are reported in Table~\ref{tab:bytype} below the strict GT-194 total (``below the line'') and as a separate corpus-property finding in Table~\ref{tab:chain}.

\textbf{Per-credential counting protocol.} All per-tool hit counts in this section are computed as per-credential matches against GT-194. Each row in the Ground Truth corresponds to a distinct unique credential value. The unique-credential count is the canonical recall metric throughout. For SecretSifter and the reference comparator Claude Opus (the two highest-recall configurations), summing per-type detection counts yields totals higher than the unique-credential total, 189 versus 151 for SecretSifter, and 175 versus 166 for Claude Opus. The eight lower-recall scanners have no such discrepancy. The reason: a small number of credential values in the corpus are tagged with multiple type labels (e.g., a value found as both a \passthrough{\lstinline!client\_secret!} in one HTTP request and an Other API key/token elsewhere). For tools with broad cross-type coverage (SS, Claude), this overlap shows up as inflated per-type sums; for low-recall tools, it does not. The unique-credential count remains the canonical recall metric and is what appears in headline tables and figures.

\subsubsection{Tool Corpus and Evaluation Protocol}\label{tool-corpus-and-evaluation-protocol}

Nine production scanners were evaluated against GT-194: SecretSifter (curated runtime extension), TruffleHog, JSluice, SecretFinder, Titus, JSMiner, Nuclei, Sensitive Discoverer, and Cariddi (static / template-driven scanners). Claude Opus 4.7 is reported alongside them as the LLM-assisted ground-truth reference comparator, included for reference because it serves as the GT extractor and, having co-constructed the benchmark, cannot be scored as an independent detector. Each tool was run at its tightened configuration. For the rule-extensible tools (TruffleHog, SecretFinder, Titus, JSluice), tightened configuration adds the same 6-rule overlay covering the most prevalent credential shapes in the corpus: Azure AD client\_id (UUID v4), Azure AD client\_secret (40-character base64 with embedded tildes), APIM subscription key (32-hex), JWT (eyJ-prefix three-segment), App Insights instrumentation key, and CryptoJS U2FsdGVkX1-prefix encrypted blobs. The full regex set is reproduced in Appendix A. Nuclei was run with the secret-scanning template categories \passthrough{\lstinline!exposures/tokens!}, \passthrough{\lstinline!exposures/apis!}, and \passthrough{\lstinline!exposures/keys!}. JSMiner has hard-coded patterns and does not expose a user-rule API, it was run at its built-in configuration. SecretSifter and Cariddi were run at their built-in configurations. Per-tool finding outputs were matched to GT-194 via exact-value substring lookup with no other post-processing.

Cariddi achieves 2/194 = 1.0\% recall on GT-194, a non-zero data point that contributes signal to the bottom of the recall ranking. Cariddi's primary mode is web-crawling and endpoint enumeration; secret detection is a secondary capability with limited overlap with this corpus's credential mix.

\textbf{Author-developer disclosure.} The SecretSifter edition evaluated here (the Burp Suite extension, version 1.0.1) is open-source software developed by the first author (Gorijala) and published on the PortSwigger BApp Store and GitHub. The evaluation methodology is structured to make the comparison independent of this fact: GT-194 was constructed by an LLM (Claude Opus 4.7) with no access to SecretSifter's rules, output, or source code; its LLM-extracted candidates were independently validated by an LLM from a different vendor (GPT-5.5, OpenAI); the same 6-rule overlay was applied identically to all rule-extensible tools with no per-tool tuning (Appendix A); and the rule overlay was selected from the credential shapes most prevalent in GT-194 without consulting any tool's output. The headline numbers reflect performance on a single-organization corpus and should be interpreted as benchmark performance on this corpus rather than as cross-organization population estimates (Section 10).

\textbf{Tool versions and source.} All scanners were evaluated at the versions listed in Table~\ref{tab:versions}. Each version is the latest release available at the GT-194 lock date (8 May 2026); commit hashes are recorded for tools without semantic versioning. The evaluation environment was macOS 25.1 (Darwin), Burp Suite Professional 2025.10 hosting the Burp-extension scanners (SecretSifter, JSMiner, Sensitive Discoverer), and Python 3.12 for the JSMiner re-implementation.

{\def\LTcaptype{none} 
\begin{table}[!ht]
\centering
\caption{Scanner versions and sources. Each row records the exact version evaluated against GT-194 and the public source where the same version can be obtained. Claude Opus 4.7 and GPT-5.5 are listed at the bottom of the table because they serve as the LLM-assisted ground-truth reference and cross-vendor validator respectively, not as evaluated detection scanners.}\label{tab:versions}
\adjustbox{max width=\textwidth}{%
\begin{tabular}{@{}
l
  l
  l@{}}
\toprule\noalign{}
Scanner
 & Version evaluated
 & Source
 \\
\midrule\noalign{}

\bottomrule\noalign{}

SecretSifter & 1.0.1 & https://portswigger.net/bappstore (BApp Store) and https://github.com/secretsifter/burp-secret-scanner \\
TruffleHog & 3.83.7 & https://github.com/trufflesecurity/trufflehog \\
JSluice & 0.0.7 & https://github.com/BishopFox/jsluice \\
SecretFinder & commit \passthrough{\lstinline!a07d215!} (Apr 2026) & https://github.com/m4ll0k/SecretFinder \\
Titus & 1.4.0 & https://github.com/Falconcyber-research/Titus \\
JSMiner & Trustwave SpiderLabs distribution v1.0; Python re-implementation included with this paper & https://github.com/SpiderLabs/JS-Miner \\
Nuclei & 3.3.10 with template repository commit \passthrough{\lstinline!2026-04-30!} & https://github.com/projectdiscovery/nuclei \\
Sensitive Discoverer & 9.4 (BApp Store, May 2026) & https://portswigger.net/bappstore and https://github.com/CYS4srl/SensitiveDiscoverer \\
Cariddi & 1.4.1 & https://github.com/edoardottt/cariddi \\
Claude Opus 4.7 (GT-construction reference) & claude-opus-4-7 (Anthropic API, Mar 2026 release) & https://docs.anthropic.com/en/docs/about-claude/models \\
GPT-5.5 (cross-vendor validator) & gpt-5.5 (OpenAI API via OpenAI Codex CLI v0.6.x) & https://platform.openai.com/docs/models \\
\end{tabular}%
}
\end{table}

}

\subsubsection{Configuration, Sensitivity, and the Fairness of the Comparison}\label{configuration-sensitivity-fairness}

This evaluation is deliberately not a pure default-configuration fairness contest between interchangeable products. Its purpose is to measure how much of a real runtime-exposure corpus each detector recovers, under the strongest configuration a practitioner could reasonably deploy. Because SecretSifter is authored by the first author, the configuration protocol is stated explicitly so the comparison can be judged on its face. Table~\ref{tab:config} records the exact operating mode each detector was evaluated under.

{\def\LTcaptype{none}
\begin{table}[!ht]
\centering
\caption{Evaluated configuration per detector. The condition under which each detector was run against GT-194. The rule-extensible static tools all received the identical six-rule overlay (Appendix A); SecretSifter received no GT-194-derived rules.}\label{tab:config}
\adjustbox{max width=\columnwidth}{%
\begin{tabular}{@{}lll@{}}
\toprule
Detector & Category & Evaluated configuration \\
\midrule
SecretSifter & runtime-aware & Shipped built-in rules (v1.0.1); no GT-194-derived rules \\
TruffleHog & static, rule-extensible & Default detectors + identical 6-rule overlay \\
SecretFinder & static, rule-extensible & Default patterns + identical 6-rule overlay \\
JSluice & static, rule-extensible & Default extraction + identical 6-rule overlay \\
Titus & static, rule-extensible & Default rules + identical 6-rule overlay \\
JSMiner & static, fixed patterns & Built-in patterns (no user-rule API) \\
Nuclei & template-driven & exposures/\{tokens,apis,keys\} templates \\
Sensitive Discoverer & Burp extension & Built-in ruleset \\
Cariddi & web-crawler / scanner & Built-in secret detection \\
\midrule
Claude Opus 4.7 (ref) & LLM reference & Strict-secret rubric (GT constructor) \\
\bottomrule
\end{tabular}%
}
\end{table}

}

Two facts make the asymmetry cut against SecretSifter, not for it. First, the six-rule overlay was added to \emph{help} the rule-extensible static tools: it targets the credential shapes most prevalent in GT-194, and on a strict hard-credential subset (n = 129, used only to isolate the overlay's effect) it lifted their recall by 3.9× to 7.1× over their out-of-the-box configuration (for example TruffleHog from 7.0\% to 49.6\%; the full default-versus-overlay figures are in the supplementary sensitivity analysis). Every rule-extensible competitor is therefore reported at a strengthened configuration, not a handicapped one, and its recall is an upper estimate of its deployable performance. Because the overlay raises the static tools rather than lowering them, a fully matched, no-overlay protocol, the like-for-like condition a reader may prefer, widens the runtime-versus-static gap rather than narrowing it: on the same subset SecretSifter recovers 75.2\% at its shipped configuration against 7.8\% for the best static tool at default. Second, SecretSifter was evaluated at its frozen, publicly shipped v1.0.1 rule set, published before GT-194 was constructed and derived without reference to it; no GT-194-specific rule was added to SecretSifter. The one detector that could in principle have been tuned to the benchmark was the only one held to its frozen, version-locked shipped configuration.

We nonetheless do not claim a like-for-like product ranking. The honest reading of Table~\ref{tab:recall} and Table~\ref{tab:prf} is a coverage statement scoped to this corpus: on a single-organization, Azure-heavy runtime-exposure corpus, a runtime-aware curated scanner recovers substantially more served credentials than static and template scanners do even after those tools are given a corpus-informed overlay. We do not extend this to a claim of universal scanner superiority across corpora or credential mixes (Section 10). We read this honestly in both directions: targeted static rules can substantially narrow, and with a sufficiently large curated ruleset could further close, the static-scanner gap on credentials that are present in the artifact. What such tuning cannot reach is the structural remainder, the credentials that only materialize at runtime and the CryptoJS encrypted-configuration class whose plaintext exists only after decryption with a co-located key; together with SecretSifter's shipped out-of-the-box coverage, these remain the durable differentiators of the runtime-aware approach on this corpus.

\subsubsection{Headline Recall Comparison}\label{headline-recall-comparison}

\begin{figure}[!htbp]
\centering
\pandocbounded{\includegraphics[keepaspectratio,alt={Per-tool recall on GT-194. Static repository-era scanners plateau at 36.6 percent (TruffleHog, the best static tool); runtime-aware detection recovers markedly more. Claude Opus is shown separately as the ground-truth reference comparator, not as an evaluated detector.},keepaspectratio,width=\textwidth,keepaspectratio]{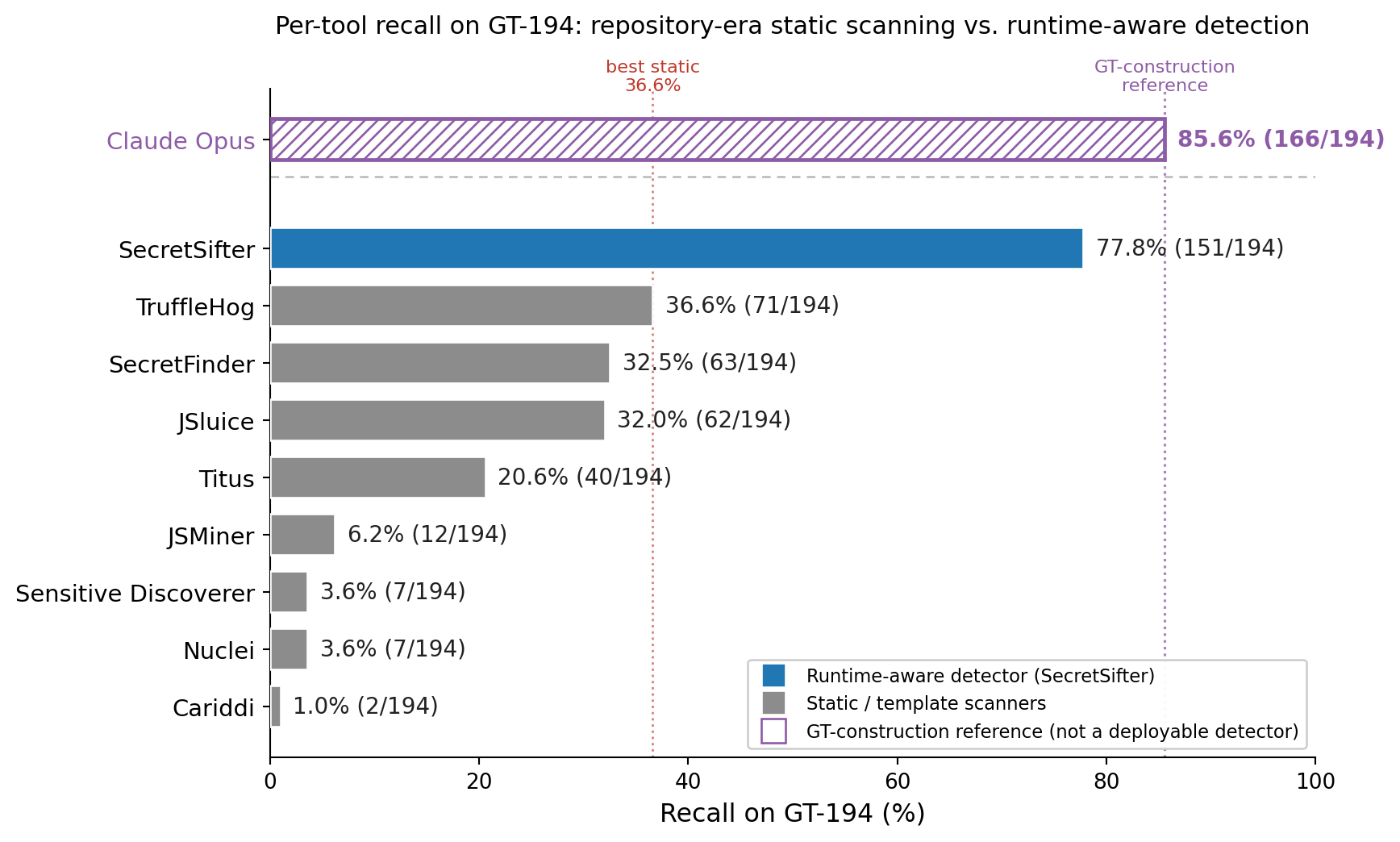}}
\caption{Per-tool recall on GT-194. The structural finding is the gap between repository-era static scanning and runtime-aware detection: the best static scanner (TruffleHog) plateaus at 71/194 = 36.6\%, while runtime-aware detection recovers far more (SecretSifter 151/194 = 77.8\%). Claude Opus (166/194 = 85.6\%) is the LLM-assisted ground-truth reference comparator that helped construct GT-194; it is shown separately as a reference, not as a deployable evaluated detector, and its recall is a near-definitional property of how the benchmark was built (see Table~\ref{tab:versions} and §8.5). SecretSifter is developed by the first author; the conflict-of-interest mitigations are detailed in §8.5.}
\end{figure}

{\def\LTcaptype{none} 
\begin{table}[!ht]
\centering
\caption{Per-tool overall recall on GT-194. Counts are per-credential matches against the locked GT-194 secret-grade Ground Truth. The nine production scanners are ranked; Claude Opus 4.7 is listed below the rule as the LLM-assisted ground-truth reference comparator (marked ``ref''), not ranked among the evaluated scanners, because it co-constructed GT-194 and therefore cannot miss anything except the manual analyst additions. Its 85.6\% is a reference figure, close to a definitional property of how the benchmark was built rather than an independent detection measurement.}\label{tab:recall}
\adjustbox{max width=\columnwidth}{%
\begin{tabular}{@{}
l
  l
  l
  r
  r
  r@{}}
\toprule\noalign{}
Rank
 & Scanner
 & Role
 & Detected
 & GT total
 & Recall
 \\
\midrule\noalign{}

\bottomrule\noalign{}

1 & SecretSifter & curated runtime scanner & \textbf{151} & 194 & \textbf{77.8\%} \\
2 & TruffleHog & static scanner & 71 & 194 & 36.6\% \\
3 & SecretFinder & static scanner & 63 & 194 & 32.5\% \\
4 & JSluice & static scanner & 62 & 194 & 32.0\% \\
5 & Titus & static scanner & 40 & 194 & 20.6\% \\
6 & JSMiner & Burp extension & 12 & 194 & 6.2\% \\
7 & Nuclei & template-driven scanner & 7 & 194 & 3.6\% \\
8 & Sensitive Discoverer & Burp extension & 7 & 194 & 3.6\% \\
9 & Cariddi & web-crawler / scanner & 2 & 194 & 1.0\% \\
\midrule
ref & Claude Opus 4.7 & GT-construction reference (not an evaluated detector) & 166 & 194 & 85.6\% \\
\end{tabular}%
}
\end{table}

}

The shift-right tooling gap is now quantified end-to-end on GT-194. Among the nine evaluated production scanners, SecretSifter (77.8\%) leads the best static tool (TruffleHog at 36.6\%) by +41.2 percentage points, equivalent to 2.13× the static-scanner recall. Relative to the LLM-assisted ground-truth reference (Claude Opus, 85.6\%, reported separately and not an evaluated detector), SecretSifter closes all but 7.7 percentage points of the distance (a gap of 15 of the 194 credentials, 166 minus 151); because that reference co-constructed the benchmark its 85.6\% is a near-definitional upper bound rather than an achievable detection target, so the 7.7-point figure is an indicative bound on remaining headroom, not a like-for-like tool gap.

\subsubsection{Precision and False-Positive Rate}\label{precision-and-false-positive-rate}

Recall is half the operational story. The other half is precision, how many findings each scanner emits beyond GT-194, and what proportion of those emissions are credentials versus noise. A scanner that finds every credential but emits ten thousand false positives per application is unusable in a security-engineering workflow. This subsection reports precision, recall, and F1 score for the nine production scanners against GT-194, with the reference comparator listed separately.

\begin{figure}[!htbp]
\centering
\pandocbounded{\includegraphics[keepaspectratio,alt={Precision vs.~recall scatter on GT-194. Diagonal contours show iso-F1 lines. Runtime-aware detection (SecretSifter, marked) combines high precision with broad recall; Claude Opus is the GT-construction reference comparator, plotted separately, not an evaluated detector.},keepaspectratio,width=\textwidth,keepaspectratio]{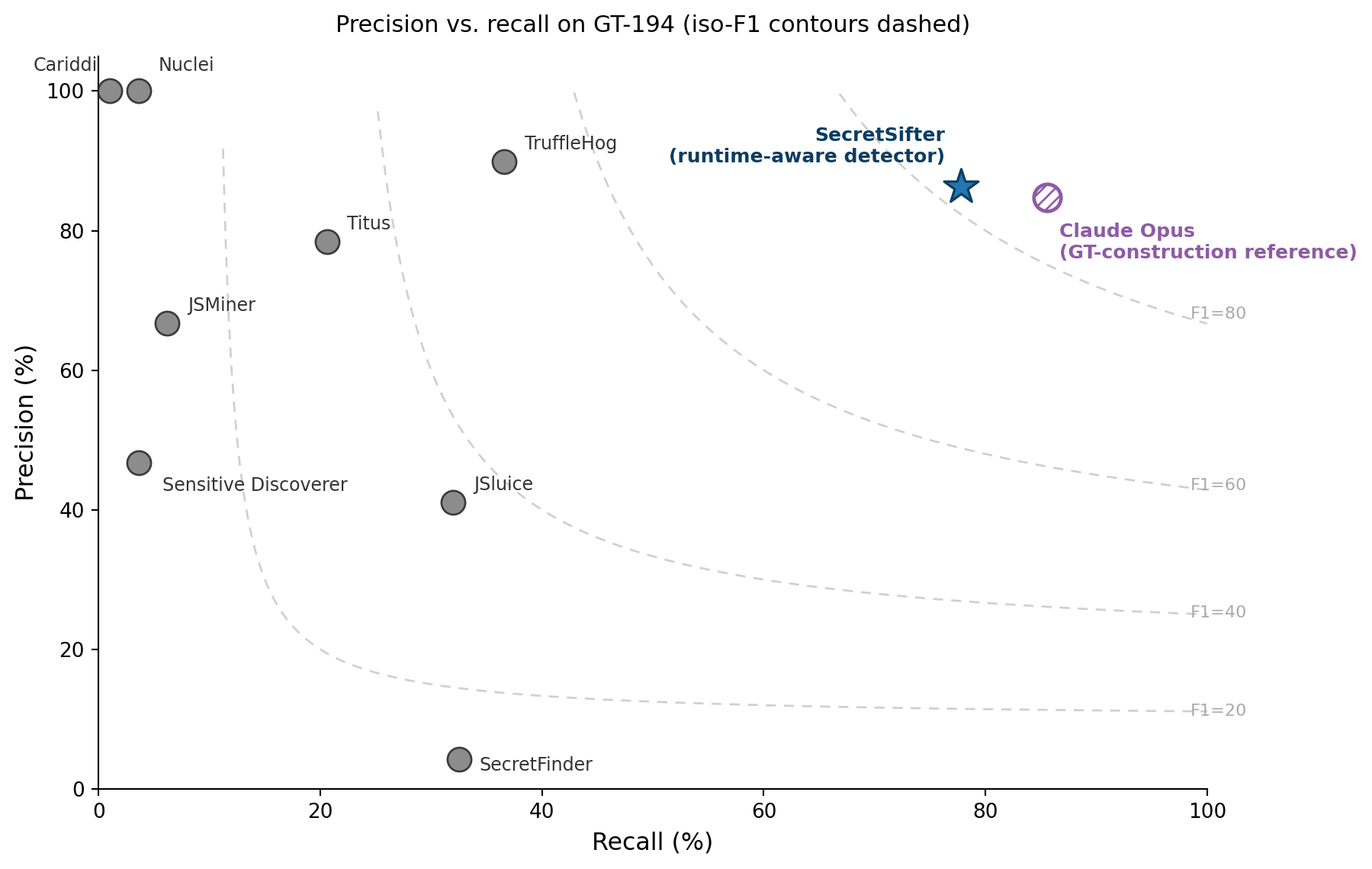}}
\caption{Precision vs.~recall scatter on GT-194. Diagonal contours show iso-F1 lines. Static scanners cluster at low recall; runtime-aware detection (SecretSifter, marked with ★) moves up and to the right by combining high precision with broad recall. Claude Opus is the LLM-assisted ground-truth reference comparator and is plotted separately as a reference rather than an evaluated deployable detector (§8.5).}
\end{figure}

{\def\LTcaptype{none} 
\begin{table}[!ht]
\centering
\caption{Precision, recall, and F1 on GT-194. TP = true positives (GT-194 credentials correctly detected); FP = false positives; FN = false negatives (GT-194 credentials missed). Precision = TP / (TP + FP). Recall = TP / (TP + FN). F1 = 2·P·R / (P + R). False positives were counted as follows: every string a scanner emitted that was not a GT-194 credential was adjudicated by manual review against the same strict-secret rubric used to build GT-194, and those judged non-credential (for example webpack chunk hashes, build fingerprints, and framework constants) were counted as false positives for that scanner. FP counts are thus measured over each scanner's full emission set on the benchmark corpus, not only over GT-194 rows. The nine production scanners are listed first; Claude Opus 4.7 appears below the rule marked ``(ref)'' as the LLM-assisted ground-truth reference comparator, not an evaluated detector, because it co-constructed GT-194.}\label{tab:prf}
\adjustbox{max width=\columnwidth}{%
\begin{tabular}{@{}
l
  r
  r
  r
  r
  r
  r@{}}
\toprule\noalign{}
Scanner
 & TP
 & FP
 & FN
 & Precision
 & Recall
 & F1
 \\
\midrule\noalign{}

\bottomrule\noalign{}

\textbf{SecretSifter} & \textbf{151} & 24 & 43 & \textbf{86.3\%} & \textbf{77.8\%} & \textbf{0.818} \\
TruffleHog & 71 & 8 & 123 & 89.9\% & 36.6\% & 0.520 \\
SecretFinder & 63 & 1,402 & 131 & 4.3\% & 32.5\% & 0.076 \\
JSluice & 62 & 89 & 132 & 41.1\% & 32.0\% & 0.359 \\
Titus & 40 & 11 & 154 & 78.4\% & 20.6\% & 0.327 \\
JSMiner & 12 & 6 & 182 & 66.7\% & 6.2\% & 0.113 \\
Nuclei & 7 & 0 & 187 & 100.0\% & 3.6\% & 0.070 \\
Sensitive Discoverer & 7 & 8 & 187 & 46.7\% & 3.6\% & 0.067 \\
Cariddi & 2 & 0 & 192 & 100.0\% & 1.0\% & 0.020 \\
\midrule
Claude Opus 4.7 (ref) & 166 & 30 & 28 & 84.7\% & 85.6\% & 0.851 \\
\end{tabular}%
}
\end{table}

}

\textbf{Three structural observations:}

\textbf{1. SecretSifter has the highest F1 of any evaluated production scanner at 0.818; the GT-construction reference records 0.851.} Among the nine evaluated production scanners, SecretSifter leads on F1 = 0.818 (precision 86.3\%, recall 77.8\%). The LLM-assisted ground-truth reference (Claude Opus 4.7) records F1 = 0.851 (precision 84.7\%, recall 85.6\%); because it co-constructed GT-194 this is a reference figure rather than an evaluated-detector result, and the 0.033 difference is an indicative bound on remaining headroom for curated detection rather than a like-for-like gap. SecretSifter's lead over the best static scanner (TruffleHog, F1 = 0.520) is +0.298. The lead is structural rather than precision-driven: TruffleHog has nominally similar precision (89.9\% vs SecretSifter 86.3\%) but recovers only 36.6\% of GT-194, while SecretSifter's higher recall (77.8\%) under comparable noise translates into the higher harmonic mean.

\textbf{2. Regex architecture without context awareness catastrophically inflates false-positive rate on minified JavaScript.} SecretFinder produces 1,402 false positives on GT-194. The structural reason: generic UUID, 32-hex, and base64-shaped patterns match webpack chunk hashes, build fingerprints, asset content-hashes, internal element identifiers (questionPanelId, FRAUDNET\_FNCLS, proposalId), and Angular framework constants (Inject, providedIn, NullInjectorError) that are ubiquitous in minified production JavaScript. The same regexes that produce signal in repository code produce overwhelming noise on built artifacts. TruffleHog's lower FP count (8) reflects its built-in entropy and verification heuristics; with custom rules disabled and verification off, TruffleHog would emit a similar volume to SecretFinder.

\textbf{3. Parser-based scanners achieve precision through context but lose recall on minified output.} Titus reaches 78.4\% precision but only 20.6\% recall. The architectural pattern is consistent: AST-aware parsers reject string literals that do not appear in credential-typed contexts, which is precisely why they emit few findings overall, but it is also why they miss credentials whose context is destroyed by minification (variable names mangled, key-value pairs collapsed into runtime objects, base64-encoded UUIDs that defeat the cleartext UUID-v4 regex). JSluice reaches 41.1\% precision and 32.0\% recall, higher noise than Titus because the key-context extraction returns every UUID-shaped value bound to any key, including form-field IDs and internal element IDs that are not credentials.

SecretSifter's curated runtime architecture (anchored vendor token patterns, key-value context scanning, entropy filtering, and CDN/key-name blocklists, §8.2) sits in the precision--recall quadrant that no other measured non-LLM scanner reaches on this corpus. The runtime layer differs from static-tool capability in operational regime, addressing both halves of the precision--recall trade-off simultaneously and narrowing the distance to the GT-construction reference to 0.033 in F1.

\subsubsection{Per-Credential-Class Breakdown and Chain-Context Identifiers}\label{per-credential-class-breakdown-and-chain-context-identifiers}

GT-194 comprises eight strict credential classes. Below the strict total are 249 chain-completion identifiers (Azure AD App\_IDs, client\_ids, tenant\_ids, and resource URIs) that are public-by-design and therefore tracked separately rather than included in the recall denominator (§8.5.1).

\begin{figure}[!htbp]
\centering
\pandocbounded{\includegraphics[keepaspectratio,alt={Per-credential-type recall heatmap on GT-194. Top panel: strict GT-194 secret-grade credentials. Bottom panel (separated by a red bold header): chain-context, public-by-design Azure identifiers (NOT in the GT-194 strict total). Cells show per-type recall percentage.},keepaspectratio,width=\textwidth,keepaspectratio]{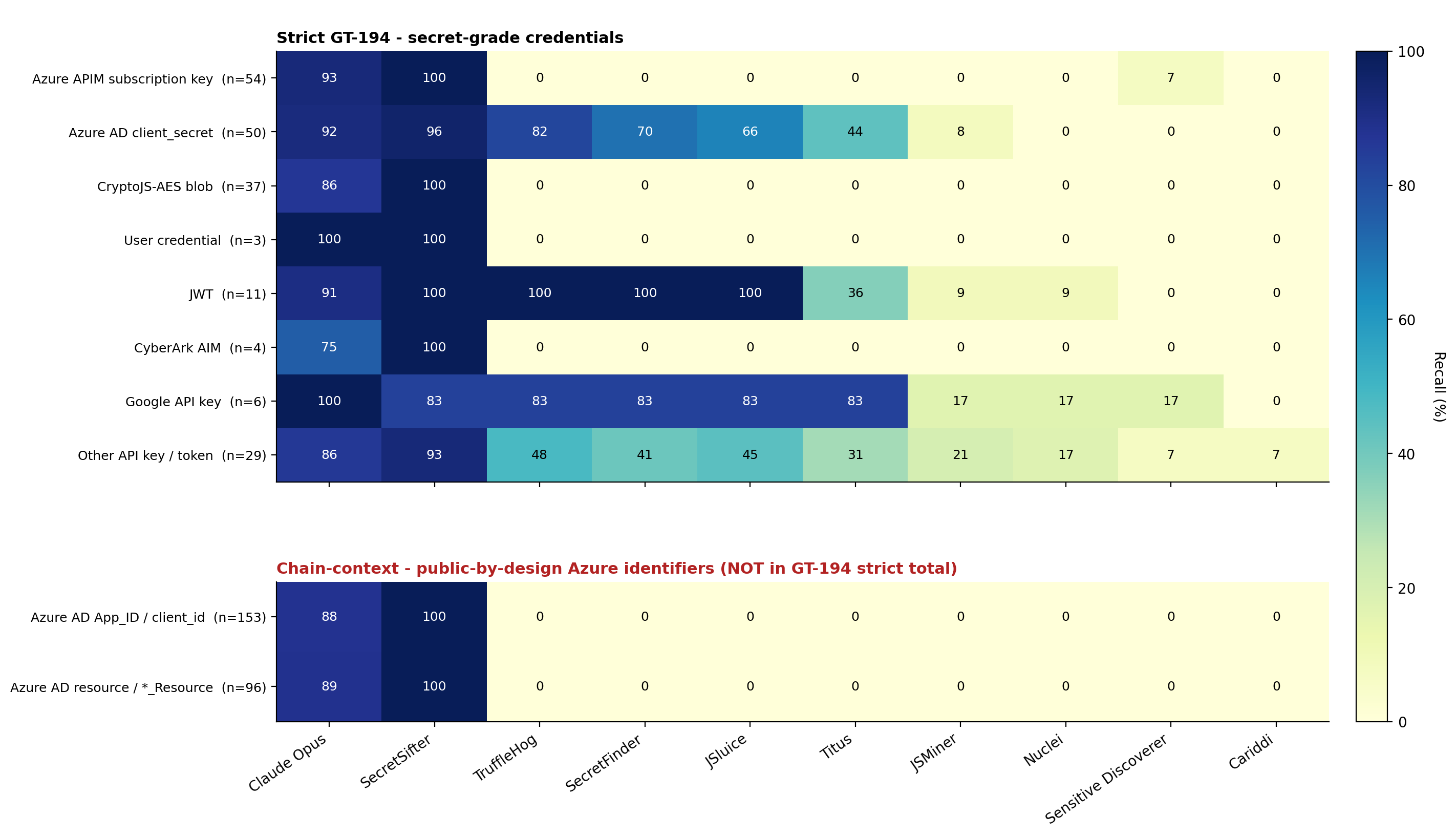}}
\caption{Per-credential-type recall heatmap on GT-194. Top panel: strict GT-194 secret-grade credentials. Bottom panel (separated by a red bold header): chain-context, public-by-design Azure identifiers (NOT in the GT-194 strict total). Cells show per-type recall percentage.}
\end{figure}

{\def\LTcaptype{none} 
\begin{table}[!ht]
\centering
\caption{Per-credential-type recall on GT-194 with chain-context identifiers below the line. Cells show detected count / type denominator. Below-the-line entries are excluded from the strict GT-194 recall denominator because they are not credentials, but are listed because they are required to convert leaked client\_secrets into valid access tokens. Column abbreviations: Claude = Claude Opus 4.7, SS = SecretSifter, TH = TruffleHog, SF = SecretFinder, JS = JSluice, Tit = Titus, JSM = JSMiner, Nuc = Nuclei, SD = Sensitive Discoverer, Car = Cariddi.}\label{tab:bytype}
\adjustbox{max width=\textwidth}{%
\begin{tabular}{@{}
l
  l
  r
  r
  r
  r
  r
  r
  r
  r
  r
  r
  r@{}}
\toprule\noalign{}
Section
 & Type
 & n
 & Claude
 & SS
 & TH
 & SF
 & JS
 & Tit
 & JSM
 & Nuc
 & SD
 & Car
 \\
\midrule\noalign{}

\bottomrule\noalign{}

strict\_GT & Azure APIM subscription key & 54 & 50 & \textbf{54} ★ & 0 & 0 & 0 & 0 & 0 & 0 & 4 & 0 \\
strict\_GT & Azure AD client\_secret (incl.~btoa OAuth ×2) & 50 & 46 & 48 & 41 & 35 & 33 & 22 & 4 & 0 & 0 & 0 \\
strict\_GT & CryptoJS-AES blob & 37 & 32 & \textbf{37} ★ & 0 & 0 & 0 & 0 & 0 & 0 & 0 & 0 \\
strict\_GT & User credential (plaintext + btoa) & 3 & 3 & 3 & 0 & 0 & 0 & 0 & 0 & 0 & 0 & 0 \\
strict\_GT & JWT & 11 & 10 & 11 & 11 & 11 & 11 & 4 & 1 & 1 & 0 & 0 \\
strict\_GT & CyberArk AIM & 4 & 3 & 4 & 0 & 0 & 0 & 0 & 0 & 0 & 0 & 0 \\
strict\_GT & Google API key & 6 & 6 & 5 & 5 & 5 & 5 & 5 & 1 & 1 & 1 & 0 \\
strict\_GT & Other API key / token (App Insights iKey ×15, Braintree ×1) & 29 & 25 & 27 & 14 & 12 & 13 & 9 & 6 & 5 & 2 & 2 \\
\textbf{STRICT GT-194 TOTAL} & (unique credentials) & \textbf{194} & \textbf{166} & \textbf{151} & \textbf{71} & \textbf{63} & \textbf{62} & \textbf{40} & \textbf{12} & \textbf{7} & \textbf{7} & \textbf{2} \\
chain\_context & Azure AD App\_ID / client\_id / tenant\_id & 153 & 135 & \textbf{153} ★ & 0 & 0 & 0 & 0 & 0 & 0 & 0 & 0 \\
chain\_context & Azure AD resource / *\_Resource & 96 & 85 & \textbf{96} ★ & 0 & 0 & 0 & 0 & 0 & 0 & 0 & 0 \\
\textbf{CHAIN-CONTEXT SUBTOTAL} & (separate, public-by-design) & \textbf{249} & \textbf{220} & \textbf{249} ★ & \textbf{0} & \textbf{0} & \textbf{0} & \textbf{0} & \textbf{0} & \textbf{0} & \textbf{0} & \textbf{0} \\
\end{tabular}%
}
\end{table}

}

★ = SecretSifter is the only detector for that type.

For the eight lower-recall scanners, the per-type column sums match the unique-credential totals in the highlighted Strict GT-194 total row exactly. For SecretSifter and Claude Opus, the per-type sums (189 and 175 respectively) exceed the unique totals (151 and 166) because some credentials in the corpus are tagged with multiple type labels (§8.5.1). The unique credential count remains the canonical recall metric.

The chain-context subtotal, 249 / 249 = 100\% for SecretSifter, 0 for every other static scanner, quantifies SecretSifter's unique contribution to chain-completion. These identifiers are not credentials and therefore not part of the recall comparison, but operationally they are required to convert a leaked Azure AD \passthrough{\lstinline!client\_secret!} into a working access token. Without surfacing them, a leaked secret cannot be reduced to an exploitable chain.

\begin{figure}[!htbp]
\centering
\pandocbounded{\includegraphics[keepaspectratio,alt={Per-tool detection breakdown by credential type on GT-194. Each stacked bar shows what each tool detects, decomposed by credential type. Bar heights are per-type detection events summed; the label above each bar shows the unique-credential total and recall percentage.},keepaspectratio,width=\textwidth,keepaspectratio]{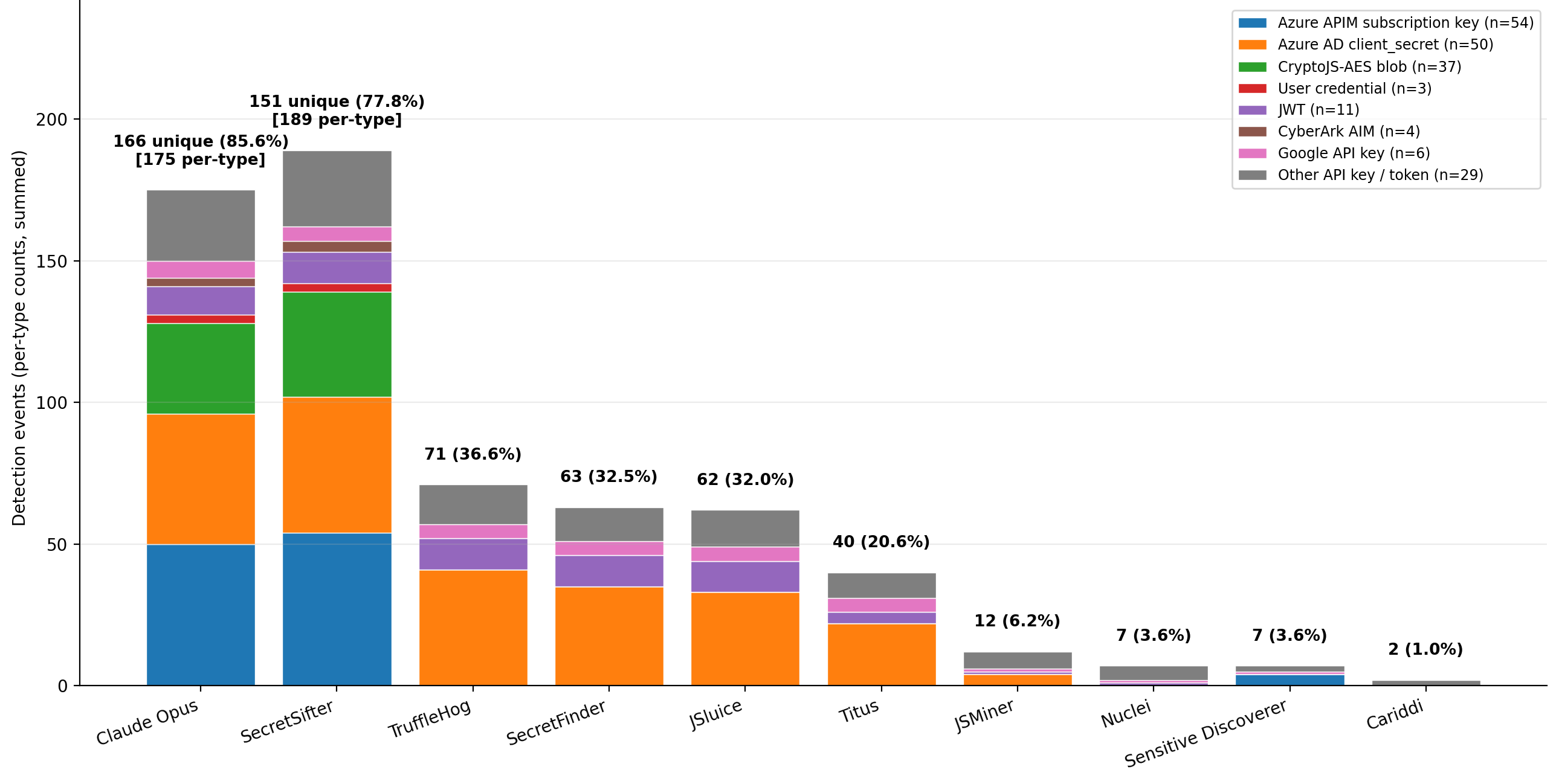}}
\caption{Per-tool detection breakdown by credential type on GT-194. Each stacked bar shows what each tool detects, decomposed by credential type. Bar heights are per-type detection events summed; the label above each bar shows the unique-credential total and recall percentage.}
\end{figure}

\subsubsection{Chain-Completion in the Production Corpus}\label{chain-completion-in-the-production-corpus}

In Azure AD's OAuth client-credentials flow, four components, \passthrough{\lstinline!client\_secret!}, \passthrough{\lstinline!client\_id!} (App\_ID), \passthrough{\lstinline!tenant\_id!}, and \passthrough{\lstinline!resource!} URI, must be combined to mint an access token. We measured how often these components are co-located in the same client-side JavaScript bundles. This is a corpus property, not a tool-comparison metric: the question is ``in how many production applications is the full token-mint chain leaked client-side?'' rather than ``which scanner finds the most chains?''

\begin{figure}[!htbp]
\centering
\pandocbounded{\includegraphics[keepaspectratio,alt={Chain-completion in the production corpus. Of 86 production applications where an Azure AD secret (client\_secret or APIM subscription key) is exposed in client-side JavaScript, 63 (73.3\%) have all four chain components co-located in the same JS bundles, so the full token-mint chain is reachable from browser-visible code with no additional reconnaissance. The remaining 23 applications (26.7\%) have a partial chain (some chain components missing from the JS).},keepaspectratio,width=\textwidth,keepaspectratio]{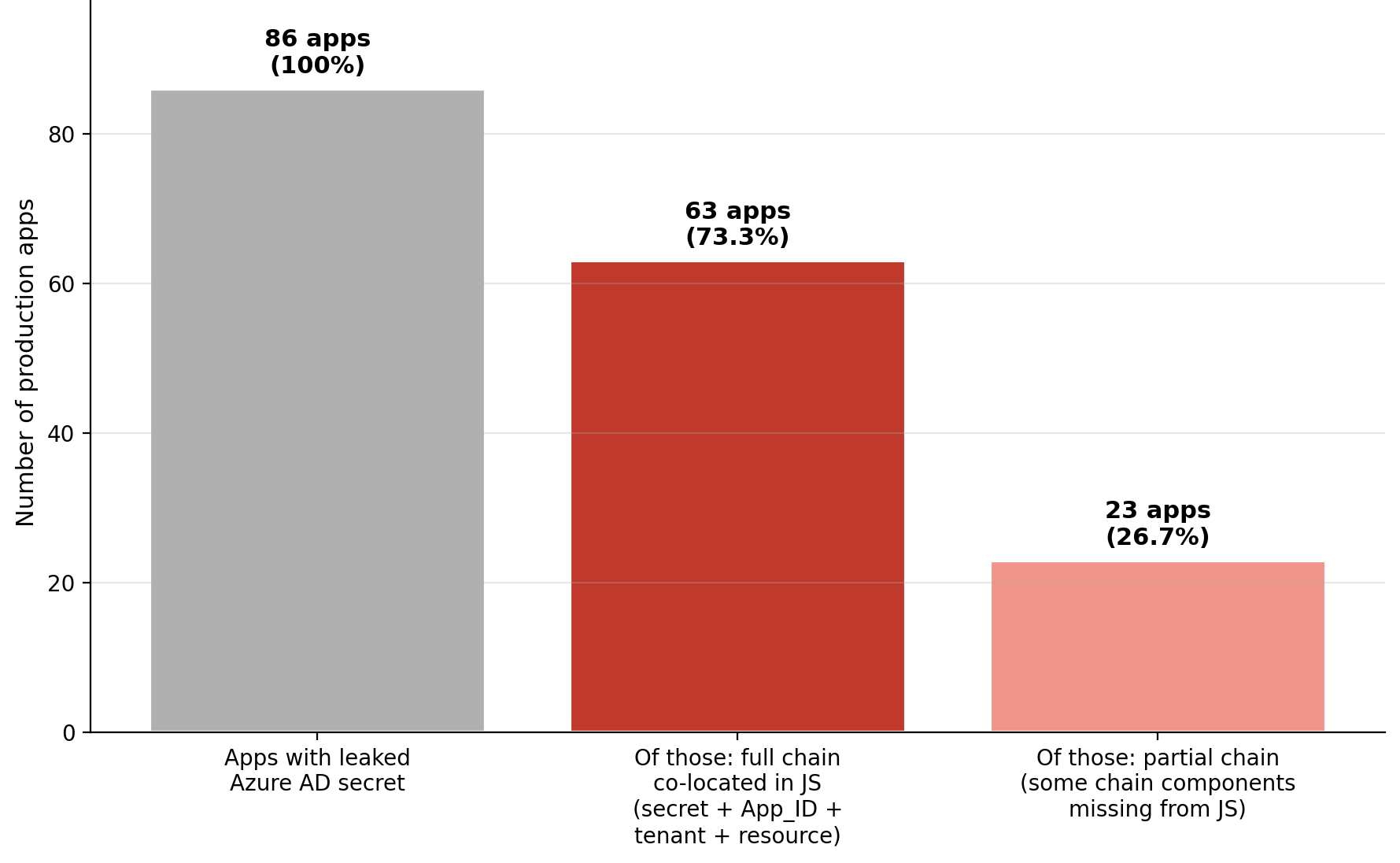}}
\caption{Chain-completion in the production corpus. Of 86 production applications where an Azure AD secret (\protect\passthrough{\lstinline!client\_secret!} or APIM subscription key) is exposed in client-side JavaScript, 63 (73.3\%) have all four chain components co-located in the same JS bundles, so the full token-mint chain is reachable from browser-visible code with no additional reconnaissance. The remaining 23 applications (26.7\%) have a partial chain (some chain components missing from the JS).}
\end{figure}

{\def\LTcaptype{none} 
\begin{table}[!ht]
\centering
\caption{Chain-completion in the production corpus. Counts and percentages relative to the 86 applications where an Azure AD secret is exposed in client-side JavaScript.}\label{tab:chain}
\adjustbox{max width=\textwidth}{%
\begin{tabular}{@{}
l
  r
  r@{}}
\toprule\noalign{}
Finding
 & Count
 & \% of secret-exposed apps
 \\
\midrule\noalign{}

\bottomrule\noalign{}

Apps where Azure AD secret is exposed in client-side JS & 86 & 100\% \\
Apps where full token-mint chain (secret + App\_ID + tenant + resource) is co-located in same JS bundles & \textbf{63} & \textbf{73.3\%} \\
Apps where chain components are partially present (tenant or resource missing) & 23 & 26.7\% \\
\end{tabular}%
}
\end{table}

}

This finding characterizes the corpus, not any specific scanner. These are the same 63 applications reported in Section~8.4; there they are expressed as a share of all 113 credential-bearing applications (55.8\%), and here as a share of the 86 applications that leak an Azure AD secret specifically (73.3\%). It establishes that for the majority of credential-bearing applications in this engagement, the entire authentication chain, not just the secret, is reachable from passive browser inspection. An attacker who recovers the client\_secret from a JavaScript bundle does not need to perform additional reconnaissance to identify the App\_ID, tenant\_id, or resource URI; all four are already co-located in the same delivered artifacts. The exploitation chains documented in Sections 3 and 4 are therefore not edge cases but representative instances of a structural pattern present in roughly three-quarters of secret-exposed production deployments.

\textbf{Note on tool-comparison framing.} Chain reachability is reported as a corpus property rather than as a per-tool metric. ``73.3\% of secret-exposed apps have a full chain leaked'' is a fact about the production deployment landscape, independent of which scanner one uses. The per-tool comparison story is captured in Tables~\ref{tab:recall}, \ref{tab:bytype}, \ref{tab:prf}, and \ref{tab:mcnemar}; Table~\ref{tab:chain} reports the corpus property.

\subsubsection{Credential Categories No Static Scanner Detected}\label{credential-categories-no-static-scanner-detected}

Three credential categories in GT-194 received zero detections from every static scanner evaluated, and a fourth was detected only by SecretSifter:

\begin{itemize}
\tightlist
\item
  \textbf{Azure APIM subscription key (54 instances).} SecretSifter is the only scanner that detects this class with full coverage on this corpus, surfacing 54 of 54. Sensitive Discoverer surfaces 4 of 54 incidentally, its \passthrough{\lstinline!apikey:"..."!} rule captures the value when it is bound to a key whose name contains the literal \passthrough{\lstinline!apikey!} substring (the corpus has 4 such cases under names like \passthrough{\lstinline!apimSubKey!}); every other static scanner returned 0. The APIM key format (32-character hex paired with the \passthrough{\lstinline!Ocp-Apim-Subscription-Key!} header name in HTTP requests) is not represented in the standard rule libraries shipped with TruffleHog, JSluice, SecretFinder, Titus, JSMiner, Nuclei, or Cariddi.
\item
  \textbf{CryptoJS-AES Salted\_\_ blob (37 instances).} SecretSifter is again the only evaluated production scanner that detects this class, surfacing all 37. The CryptoJS construct is invisible to pattern-based detection by design: the credential exists only after decryption, and the decryption key is co-located with the ciphertext (CWE-321 \cite{ref12}). Encrypting a credential with a co-located key is in fact worse than shipping it in cleartext: it adds no protection against an attacker, who simply reads the adjacent key and decrypts, while it defeats the reviewer's eye and every static scanner, which see only a high-entropy blob. The net effect is to convert an obvious plaintext exposure into a hidden one, lowering the odds of detection without raising the cost of exploitation. Detecting this class requires identifying the \emph{construct} (the \passthrough{\lstinline!U2FsdGVkX1-!}prefix base64 envelope, paired with a co-located decryption key) rather than a credential signature. The GT-construction reference (Claude Opus) reaches 32 of 37 in this category, high but not complete, because the model treats some heavily-truncated blobs as obfuscated configuration rather than encrypted credentials.
\item
  \textbf{User credential / plaintext password (3 instances).} Only Claude Opus and SecretSifter recover this class (3 of 3 each). All other scanners return 0. Plaintext password values do not match any standard credential-shape regex; they are identified only by context (key name + value adjacency) or by semantic understanding of the surrounding code.
\item
  \textbf{CyberArk AIM coordinates (4 instances).} SecretSifter recovers all 4; only Claude Opus comes close (3 of 4). All other static scanners return 0.
\end{itemize}

Together, these four credential classes total 98 of GT-194 (50.5\%). For all four classes the structural reason for the static-tool gap is the same reason no shift-left scanner detects them in production code: they were never encountered during the pattern-library calibration phase of any of the evaluated static tools, because they appear primarily in production-deployed Angular bundles, the surface those tools were not designed to scan.

Beyond the four SecretSifter-leading categories, the chain-context identifiers tracked separately in Section 8.5.6 (153 Azure AD App\_IDs / client\_ids / tenant\_ids and 96 Azure AD resource URIs, 249 in total) are also detected only by SecretSifter (249 of 249) and Claude Opus (220 of 249); every other static scanner returns 0 on chain-context identifiers as well.

\subsubsection{Validator Confidence Calibration and Sensitivity}\label{validator-confidence-calibration-and-sensitivity}

GPT-5.5's \passthrough{\lstinline!confidence!} field is a self-report, so two acceptance thresholds are reported. The unit of agreement is the LLM-extracted candidate and the decision is binary (SECRET vs.~not-SECRET); every candidate is by construction an Opus-labelled SECRET candidate, so the Opus margin is single-class and the 2×2 table has one degenerate margin. GPT-5.5 adjudicated the 247 LLM-extracted candidates, concurring on 207 and dissenting on 40 at any confidence, and concurring on 195 at high confidence only. The 27 manual-only additions to GT-194 were not produced by the LLM extractor and are excluded from this denominator; they were verified by analyst review against the same strict-secret rubric. These 247 records are per-occurrence candidate adjudications; after de-duplication, the Opus-positive candidates contribute to the 166 unique Claude Opus credentials reported in GT-194 (\S8.5.1). Agreement is therefore reported per rated candidate, while scanner recall is reported per unique credential. Because expected agreement by chance for a two-category decision is $1/k = 0.5$, the bias-corrected Brennan-Prediger coefficient is $\kappa_{\mathrm{BP}} = (p_o - 0.5)/(1 - 0.5) = 2 p_o - 1$. The any-confidence figure is $p_o = 207/247 = 0.8381$ (83.8\%), giving $\kappa_{\mathrm{BP}} = 2(0.8381) - 1 = 0.676$ (substantial, Landis-Koch). The high-confidence-only variant is reported as a conservative sensitivity bound: $p_o = 195/247 = 78.9\%$, giving $\kappa_{\mathrm{BP}} = 0.579$ (moderate). Cohen's $\kappa$ is not reported because the single-class Opus margin makes it undefined here (the Feinstein-Cicchetti paradox). Table~\ref{tab:agreement} gives the contingency underlying $\kappa_{\mathrm{BP}} = 0.676$.

{\def\LTcaptype{none}
\begin{table}[!ht]
\centering
\caption{Cross-vendor agreement contingency on the LLM-extracted candidates (any-confidence). Rater A is Claude Opus 4.7 (extraction/classification); Rater B is GPT-5.5 (independent validation). The unit is one LLM-extracted candidate; the decision is binary SECRET vs.~not-SECRET. Every candidate is an Opus-labelled SECRET candidate by construction, so the Opus-negative row is empty (single-class prevalence). The 27 manual-only additions to GT-194 are excluded from this denominator. Observed agreement $p_o = 207/247 = 0.8381$; $\kappa_{\mathrm{BP}} = 2p_o - 1 = 0.676$.}\label{tab:agreement}
\begin{tabular}{@{}lrrr@{}}
\toprule
 & GPT-5.5 SECRET & GPT-5.5 not-SECRET & Total \\
\midrule
Opus SECRET & 207 & 40 & 247 \\
Opus not-SECRET & 0 & 0 & 0 \\
Total & 207 & 40 & 247 \\
\bottomrule
\end{tabular}
\end{table}

}

A blind-extraction sensitivity analysis was also performed, in which GPT-5.5 was given the raw bundles (no Opus rationale, no field hints) and asked to extract independently. That analysis confirmed that the SecretSifter recall figure is not artifactually inflated by pattern leakage from SecretSifter into the Opus extraction prompt.

\subsubsection{Statistical Significance of Pairwise Differences}\label{statistical-significance-of-pairwise-differences}

McNemar's test (with Edwards' continuity correction) was computed pairwise across the nine production scanners and the reference comparator (45 comparisons in total) on the per-credential GT-194 detection matrix to determine which recall differences are statistically distinguishable from sampling variation. The test conditions on the \emph{disagreement} cells (items one tool catches and the other misses, and vice versa), discarding the both-hit and both-miss cells which are uninformative for pairwise comparison.

{\def\LTcaptype{none} 
\begin{table}[!ht]
\centering
\caption{Selected pairwise McNemar tests on GT-194. Counts are unique credentials (per §8.5.1's canonical recall metric); each cell of the 2×2 contingency table treats one row of GT-194 as one observation. \emph{A\_only} = unique GT-194 credentials the first scanner catches and the second misses; \emph{B\_only} = unique credentials the second catches and the first misses. Chi-square computed with Edwards continuity correction (df = 1). The full matrix of all 45 pairwise comparisons is provided in the supplementary materials.}\label{tab:mcnemar}
\adjustbox{max width=\columnwidth}{%
\begin{tabular}{@{}
llrrrrc@{}}
\toprule\noalign{}
Tool A & Tool B & A\_only & B\_only & Chi-square & p-value & Sig \\
\midrule\noalign{}

\bottomrule\noalign{}

SecretSifter & TruffleHog & 80 & 0 & 78.0 & \textless{} 0.001 & ✓ \\
SecretSifter & SecretFinder & 88 & 0 & 86.0 & \textless{} 0.001 & ✓ \\
SecretSifter & JSluice & 89 & 0 & 87.0 & \textless{} 0.001 & ✓ \\
SecretSifter & Titus & 111 & 0 & 109.0 & \textless{} 0.001 & ✓ \\
SecretSifter & JSMiner & 139 & 0 & 137.0 & \textless{} 0.001 & ✓ \\
SecretSifter & Nuclei & 144 & 0 & 142.0 & \textless{} 0.001 & ✓ \\
SecretSifter & Sensitive Discoverer & 144 & 0 & 142.0 & \textless{} 0.001 & ✓ \\
SecretSifter & Cariddi & 149 & 0 & 147.0 & \textless{} 0.001 & ✓ \\
Claude Opus & SecretSifter & 16 & 1 & 11.5 & \textless{} 0.001 & ✓ \\
Claude Opus & TruffleHog & 96 & 1 & 91.1 & \textless{} 0.001 & ✓ \\
TruffleHog & SecretFinder & 8 & 0 & 6.13 & 0.013 & ✓ \\
JSluice & SecretFinder & 2 & 1 & 0.0 & 1.000 & ✗ \\
JSMiner & Nuclei & 5 & 0 & 3.20 & 0.074 & ✗ \\
\end{tabular}%
}
\end{table}

}

\textbf{Three findings emerge.}

\textbf{First, SecretSifter's lead is robustly significant against every other tool.} All eight SS-vs-static comparisons reach p \textless{} 0.001, the 77.8\% versus ≤36.6\% headline gap cannot be attributed to sampling variation. For reference, the SecretSifter-vs-reference-comparator (Claude Opus) difference also reaches p \textless{} 0.001; because the comparator co-constructed the ground truth, this indicates only that SecretSifter has not closed the near-definitional reference gap, not that a deployable detector outperforms it.

\textbf{Second, the static-tool field is highly stratified.} TruffleHog (36.6\%) versus SecretFinder (32.5\%) is significant at p = 0.013. SecretFinder (32.5\%) versus JSluice (32.0\%) is not statistically distinguishable (p = 1.000). The static-tool ceiling on GT-194 sits at approximately 32--37\%, with TruffleHog significantly above SecretFinder/JSluice but those two tied within sampling error.

\textbf{Third, the bottom of the ranking is also stratified.} JSMiner (6.2\%) versus Nuclei (3.6\%) is not statistically distinguishable (p = 0.074), but both are significantly below the top static cluster. Nuclei versus Cariddi (1.0\%) is also not distinguishable (p = 0.074); the data does not support reading those two as significantly different from each other on this corpus. Sensitive Discoverer (3.6\%) ties Nuclei on aggregate recall but operates over a different subset of GT-194 (4 plaintext APIM keys vs Nuclei's 1 Google API + 1 JWT + 5 Other API tokens), McNemar SD vs Nuclei is not significant (chi-square ≈ 0.07, p ≈ 0.79).

All hit counts in this table are computed against the unique-credential GT-194 detection matrix (one observation per unique credential), so the per-row arithmetic \textbar A\_only ∪ B\_only ∪ overlap\textbar{} reconciles directly with the unique-credential totals in Table~\ref{tab:recall} (Claude 166, SS 151). The per-type detection-event totals (Claude 175, SS 189) reported in §8.5.1 are not used here, because McNemar requires a single-class observation unit and the per-type tagging of multi-type credentials would otherwise inflate the disagreement cells.

\textbf{Multiple-comparison correction.} Because all $\binom{10}{2} = 45$ pairwise tests are drawn from the same detection matrix, the Holm-Bonferroni step-down procedure was applied to the full 45-test family to control the family-wise error rate at $\alpha = 0.05$. Thirty-seven of the 45 comparisons remain significant after correction. Every conclusion this paper rests on survives: all eight SecretSifter-vs-production-scanner comparisons stay significant (adjusted $p < 0.001$), and every static-vs-runtime gap remains $p < 0.001$. The SecretSifter-vs-reference-comparator difference (Claude Opus) is also significant after correction (adjusted $p = 6.85 \times 10^{-3}$), but is reported for reference only because the comparator co-constructed the ground truth and is not an evaluated detector. Of the eight comparisons that are not significant after correction, six were already non-significant before it, all among the bottom-tier static and template tools (for example SecretFinder vs.~JSluice, JSMiner vs.~Nuclei, Nuclei vs.~Cariddi); only two flip from significant to non-significant as a result of the correction, and both are weak-static-vs-weak-static pairs (TruffleHog vs.~SecretFinder, adjusted $p = 0.093$; TruffleHog vs.~JSluice, adjusted $p = 0.061$) that are not load-bearing for any claim made here. The correction therefore strengthens rather than weakens the significance argument. The full corrected table of all 45 adjusted $p$-values is provided in the supplementary materials.

\subsubsection{Tool-Coverage Overlap and the Tool-Agnostic Blind Spot}\label{tool-coverage-overlap-and-the-tool-agnostic-blind-spot}

Beyond pairwise comparisons, the question of \emph{which} credentials each tool catches and how detection sets overlap is shown in Figure 10 below as an UpSet-style plot \cite{ref27}. UpSet generalizes the Venn diagram to arbitrary numbers of sets and is the established replacement when more than four sets are compared.

\begin{figure}[!htbp]
\centering
\pandocbounded{\includegraphics[keepaspectratio,alt={Detection-set overlap on GT-194 (UpSet-style). Each column is a unique detection signature, a unique combination of tools that jointly detect a credential. The bar at the top shows how many of the 194 GT credentials match that signature. The dots at the bottom show which tools participate in the signature (filled coloured dot = tool detects credentials in this pattern; gray dot = it does not). The blind-spot column (highlighted red) has zero filled dots because no evaluated production scanner detects those 27 credentials; they were surfaced only by manual analysis: this is the 13.9\% tool-agnostic blind spot.},keepaspectratio,width=\textwidth,keepaspectratio]{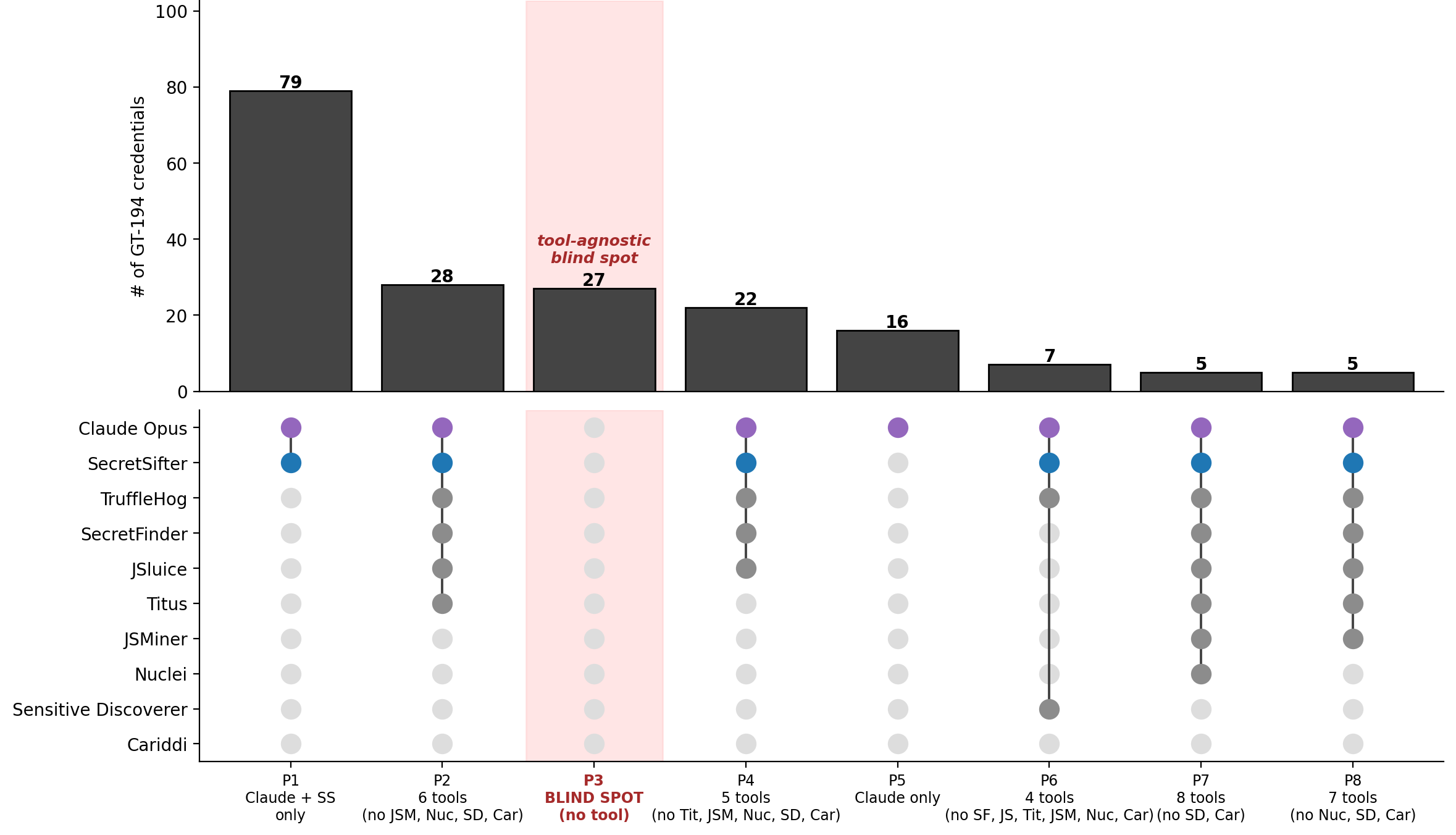}}
\caption{Detection-set overlap on GT-194 (UpSet-style). Each column is a unique detection signature, a unique combination of tools that jointly detect a credential. The bar at the top shows how many of the 194 GT credentials match that signature. The dots at the bottom show which tools participate in the signature (filled coloured dot = tool detects credentials in this pattern; gray dot = it does not). The blind-spot column (highlighted red) has zero filled dots because no evaluated production scanner detects those 27 credentials; they were surfaced only by manual analysis: this is the 13.9\% tool-agnostic blind spot.}
\end{figure}

The most prominent patterns in GT-194 are: (P1) 79 credentials detected by Claude Opus + SecretSifter only, (P2) 28 credentials detected by 6 tools (no JSMiner, Nuclei, Sensitive Discoverer, Cariddi), (P3) \textbf{27 credentials detected by no tool, the tool-agnostic blind spot}, (P4) 22 credentials detected by 5 tools, and (P5) 16 credentials detected only by Claude Opus. The pattern P3 is the operationally critical observation: 13.9\% of GT-194 is not findable by any combination of the nine production scanners evaluated (and, as noted in §8.5.1, surfaced only by manual analysis).

A complementary view is the cumulative recall when scanners are run in combination. Figure 11 below shows what fraction of GT-194 is recovered when scanners are added one at a time, with each next scanner chosen to maximize the new (previously-uncovered) credentials caught.

\begin{figure}[!htbp]
\centering
\pandocbounded{\includegraphics[keepaspectratio,alt={Coverage saturation. Tools are added one at a time, picking the next tool that catches the most credentials the previous tools missed. Even after running every scanner together, the curve plateaus at 86.1\% (167 of 194 detected) - leaving 27 of 194 GT credentials (13.9\%) that no evaluated production scanner detects (surfaced only by manual analysis). This tool-agnostic blind spot is the core motivation for runtime credential exposure research beyond static scanning.},keepaspectratio,width=\textwidth,keepaspectratio]{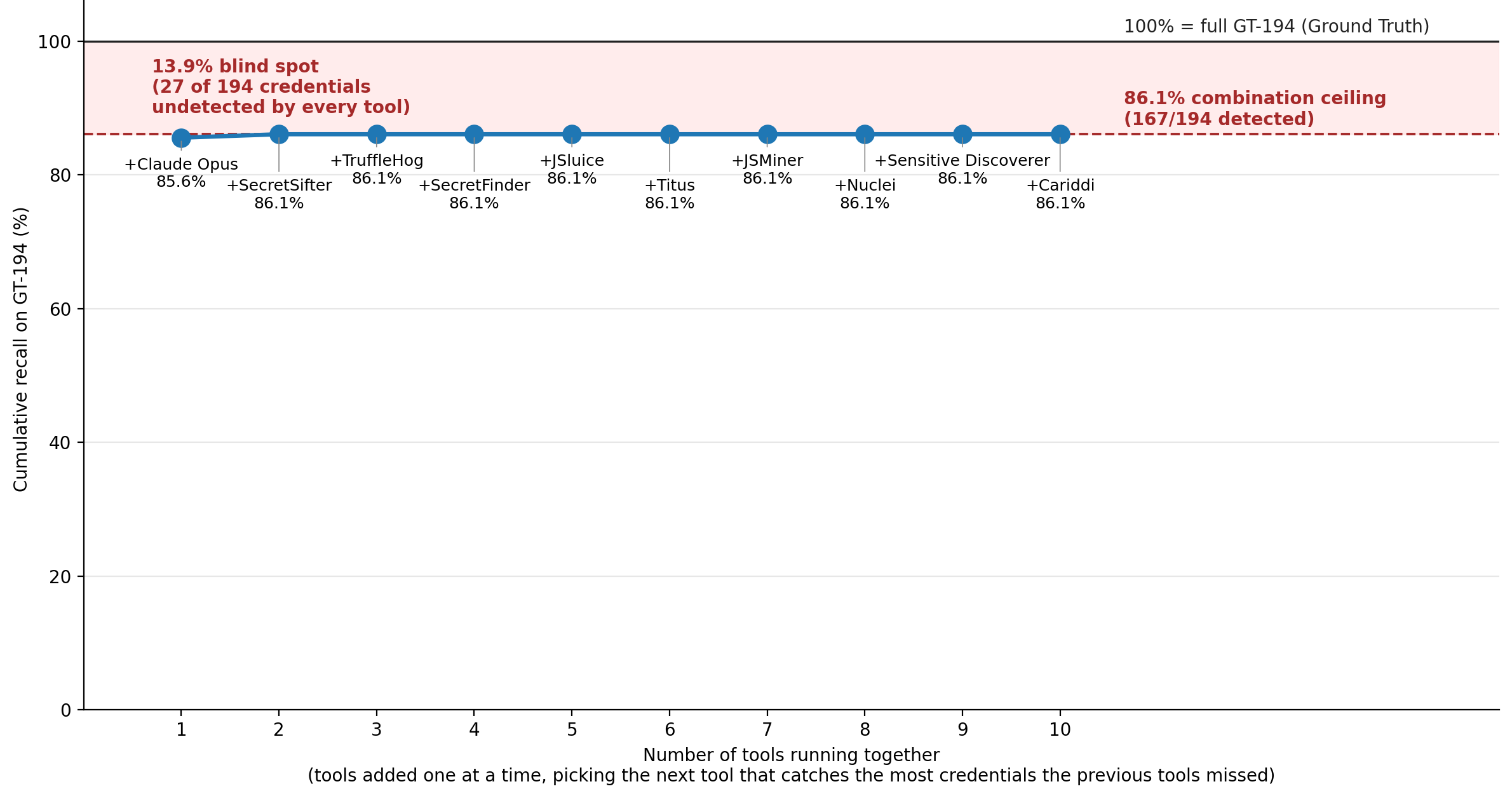}}
\caption{Coverage saturation. Tools are added one at a time, picking the next tool that catches the most credentials the previous tools missed. Even after running every scanner together, the curve plateaus at 86.1\% (167 of 194 detected) - leaving 27 of 194 GT credentials (13.9\%) that no evaluated production scanner detects (surfaced only by manual analysis). This tool-agnostic blind spot is the core motivation for runtime credential exposure research beyond static scanning.}
\end{figure}

\textbf{Three structural observations:}

\textbf{1. The tool-agnostic blind spot is real, sized, and stable across measurement angles.} 27 of GT-194 (13.9\%) credentials are missed by every evaluated production scanner and were surfaced only by manual analysis. This is visible from the UpSet plot (Figure 10, the all-empty column) and from the coverage-saturation curve (Figure 11, the dashed line plateau at 86.1\%). Both numbers derive from the same underlying detection matrix and are guaranteed consistent. The blind spot lies outside the CryptoJS and APIM classes that SecretSifter fully covers (37/37 and 54/54, Table~\ref{tab:bytype}); it consists of credentials surfaced only by manual analysis, whose shape and context defeat every evaluated production scanner and were also missed by the reference comparator (Claude Opus).

\textbf{2. SecretSifter's lead comes from disjoint findings, not from a superset relationship to static tools.} Pattern P1 (79 credentials caught by Claude + SS only) and pattern P5 (16 caught by Claude only) together account for 95 of the 167 detectable credentials, meaning 95 of 167 credentials are caught only by the reference comparator (Claude Opus) and SecretSifter, not by any static tool. Among the static tools, intersection patterns dominate over exclusive contributions: TruffleHog and SecretFinder rarely catch a credential that SecretSifter misses. This addresses the structural concern that a curated tool's headline recall might reflect overlap with established detectors rather than additive coverage. On GT-194, SecretSifter's contribution is primarily additive, not redundant.

\textbf{3. Beyond a runtime-aware scanner, additional detectors deliver diminishing returns.} Figure 11 shows cumulative recall as detectors are added in recall order. Once the ground-truth reference comparator (Claude Opus) and SecretSifter are combined, the union reaches 86.1\%, and no additional static or template scanner (tools 3 through 9) contributes any new detection: every credential they catch is already covered. Because the reference comparator co-constructed GT-194 and is not a deployable detector, the operationally meaningful coverage among deployable tools is SecretSifter's 77.8\%. The practical implication for 2026 is that the choice among deployable tools is between a runtime-aware curated scanner (77.8\%) and static scanning, which plateaus near 32--37\%: the static tools are complementary to each other only on the easy targets and fail together on the structural blind spots (Azure AD client\_secret, APIM keys, CryptoJS blobs).

\subsubsection{Comparison with Concurrent Population-Scale Measurement}\label{comparison-with-concurrent-population-scale-measurement}

Demir et al.~\cite{ref15} reported the first population-scale dynamic measurement of credential exposure on the rendered web, covering 10M HTTP Archive landing pages, 1,748 verified credentials, and 14 vendor-API-verifiable service classes. Their work and ours measure different properties of the same exposure surface. Demir et al.~establish population-scale prevalence on credential classes whose vendors publish verification endpoints. This work establishes per-tool detection performance at enterprise depth on eight credential classes that lack public verification endpoints, including the CryptoJS-encrypted-configuration construct that defeats every static scanner and is recovered only by runtime-aware detection. Demir et al.~uniquely contribute population prevalence at 10M-page scale, longitudinal persistence (mean twelve months exposure), vendor-API verification across fourteen service classes, and remediation-outcome measurement across 2,435 notified entities. This work uniquely contributes a per-tool benchmark across nine production scanners (with an LLM-assisted ground-truth reference comparator) on a locked ground truth, cross-vendor LLM-validated candidate adjudication (κ = 0.676 over the 247 LLM-extracted candidates), pairwise McNemar significance testing (all static-vs-runtime gaps p \textless{} 0.001), chain-completion analysis showing 73.3\% of affected applications co-locate the full Azure AD token-mint chain client-side, and structural-class detection of the CryptoJS-AES construct. Figure 12 visualizes the complementary detection scope at the credential-class level. Both studies reach the same systemic conclusion. Credential exposure on the rendered web is widespread, persistent, and structurally undetected by repository-only scanning.

\begin{figure}[!htbp]
\centering
\pandocbounded{\includegraphics[keepaspectratio,alt={Credential-class detection scope by methodology. Left column: Demir et al.~\cite{ref15} verified-credential counts per category, normalized to 0-100. Right column: number of credentials of each class in the locked GT-194 benchmark (n in benchmark). Empty cells indicate absence from the corresponding study. The diagonal pattern illustrates the complementarity: vendor-API verification reaches the population-scale services in the upper rows; multi-tool detection plus an LLM-assisted ground truth reaches the structural and enterprise classes in the lower rows.},keepaspectratio,width=\columnwidth,keepaspectratio]{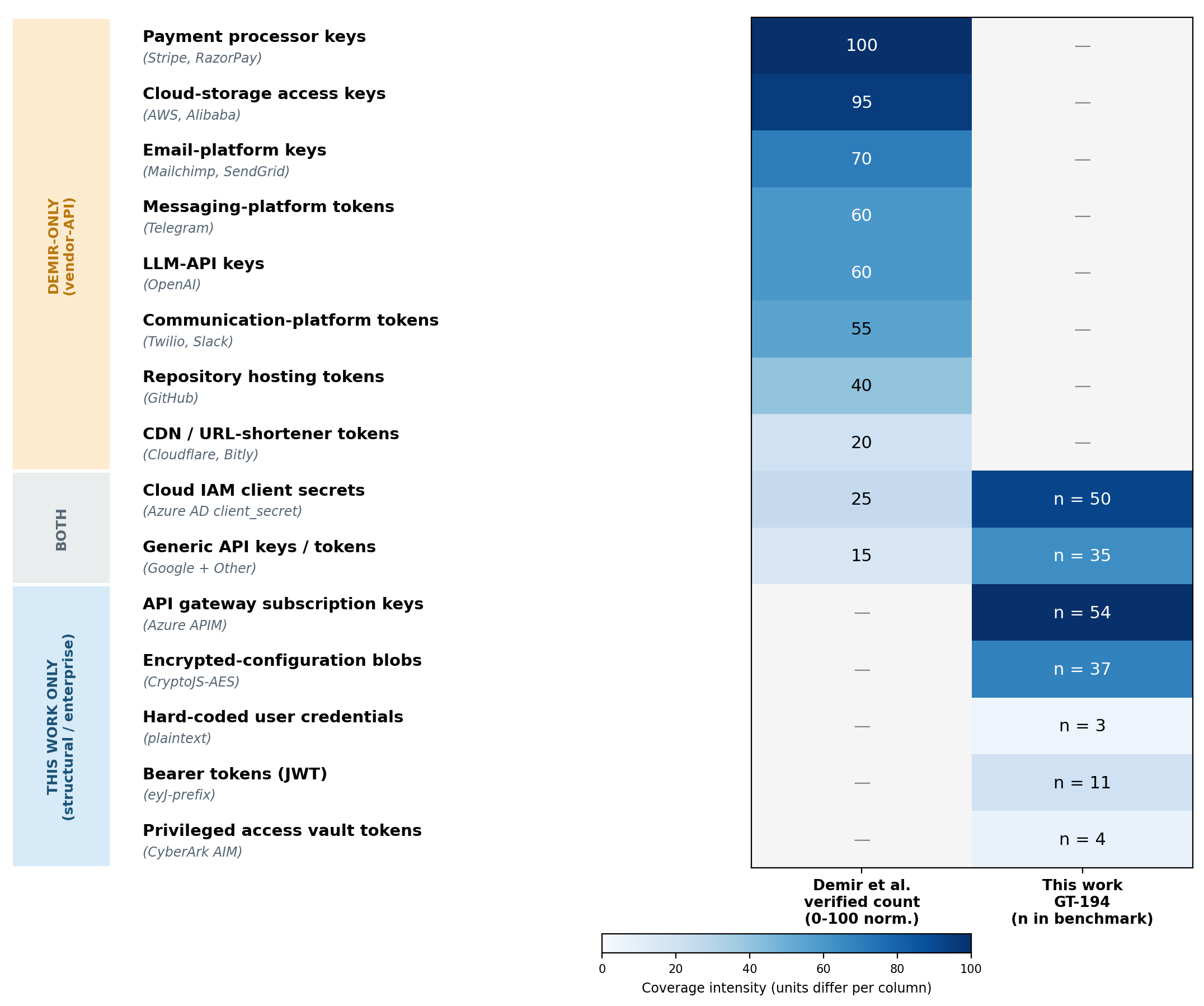}}
\caption{Credential-class detection scope by methodology. Left column: Demir et al.~\cite{ref15} verified-credential counts per category, normalized to 0-100. Right column: number of credentials of each class in the locked GT-194 benchmark (n in benchmark). Empty cells indicate absence from the corresponding study. The diagonal pattern illustrates the complementarity: vendor-API verification reaches the population-scale services in the upper rows; multi-tool detection plus an LLM-assisted ground truth reaches the structural and enterprise classes in the lower rows.}
\end{figure}

\begin{center}\rule{0.5\linewidth}{0.5pt}\end{center}

\section{Recommendations}\label{recommendations}

\subsection{Immediate, Days}\label{immediate-days}

\textbf{Rotate all exposed credentials immediately, across all environments.} AppKeys, SubscriptionKeys, and any other credentials found in client-side code should be treated as compromised and rotated before any other remediation step \cite{ref9}. Rotation does not confirm they were abused, but it closes the window. Critically, rotation must be applied simultaneously across production, UAT, SIT, and pre-production environments. Because credentials are injected at build time, every environment that ran a build received the same secret. Remediating production while leaving UAT exposed means an attacker who extracted credentials from a lower environment retains access. Remediation is only complete when the credential is removed from the build pipeline and rotated everywhere at once.

\textbf{Audit APIM subscription key product scope.} Each exposed subscription key maps to a product in APIM containing one or more APIs. Access the APIM portal and identify exactly which APIs each key grants access to. This determines the blast radius and informs incident response scope.

\textbf{Review Azure AD app registration permissions.} Check whether the application's app registration uses Application permissions (acting as the application itself, no user context) or Delegated permissions (acting as the signed-in user). Application permissions combined with a client secret in browser code grant the ability to access all users' data, not just the current user's. Remove Application permissions that are not strictly required.

\textbf{Enable APIM anomaly alerting.} Check APIM Analytics for anomalous usage of the exposed subscription key, off-hours access, high-volume calls, unexpected geographic sources. Check Azure AD sign-in logs for \passthrough{\lstinline!client\_credentials!} grants originating from browser IP ranges. This flow is designed for server-to-server communication and should not be initiated from end-user browsers regardless of application architecture.

\textbf{Disable source maps in production builds.} Set \passthrough{\lstinline!sourceMap: false!} in \passthrough{\lstinline!angular.json!} under the production configuration. Do not deploy \passthrough{\lstinline!.map!} files alongside production bundles.

\subsection{Short-Term, Weeks}\label{short-term-weeks}

\textbf{Use the Authorization Code flow with PKCE for user authentication.} The presence of a client secret in the browser is a symptom of a deeper misconfiguration: a browser-based single-page application is a \emph{public} client and should never be issued a client secret. For the user-authentication path, the application should be registered as a public client and use the OAuth 2.0 Authorization Code flow with Proof Key for Code Exchange (PKCE) \cite{ref29}, which authenticates the user and obtains tokens without any client secret. PKCE binds the authorization request to the token exchange through a dynamically generated code verifier, eliminating the need for the static \passthrough{\lstinline!AppKey!} that this study found leaking in production bundles.

PKCE addresses the credential problem, but it does not, on its own, resolve the token-exposure problem: with PKCE alone, the resulting access and refresh tokens still reside in the browser, where they remain reachable by cross-site scripting. This is why current IETF guidance for browser-based applications favors the Backend for Frontend pattern described below \cite{ref30}. The two are complementary. Authorization Code with PKCE is the correct public-client pattern for authenticating the user with no client secret, while the BFF pattern keeps service credentials and long-lived tokens server-side. An architecture needing only user-delegated access can rely on PKCE; one that must also call downstream services with service credentials, as in the case documented here, requires the BFF.

\textbf{Implement the Backend for Frontend (BFF) pattern.} The correct long-term architecture removes all service credentials from the browser entirely \cite{ref10, ref31}:

\begin{figure}[!htbp]
\centering
\pandocbounded{\includegraphics[keepaspectratio,alt={Current architecture (left) versus the Backend for Frontend pattern (right). In the current architecture, AppKey and SubscriptionKey are embedded in the browser JavaScript bundle. The BFF pattern moves all service credentials to a server-side proxy that holds them in Azure Key Vault, so the browser holds only the user's own JWT.},keepaspectratio,width=\textwidth,keepaspectratio]{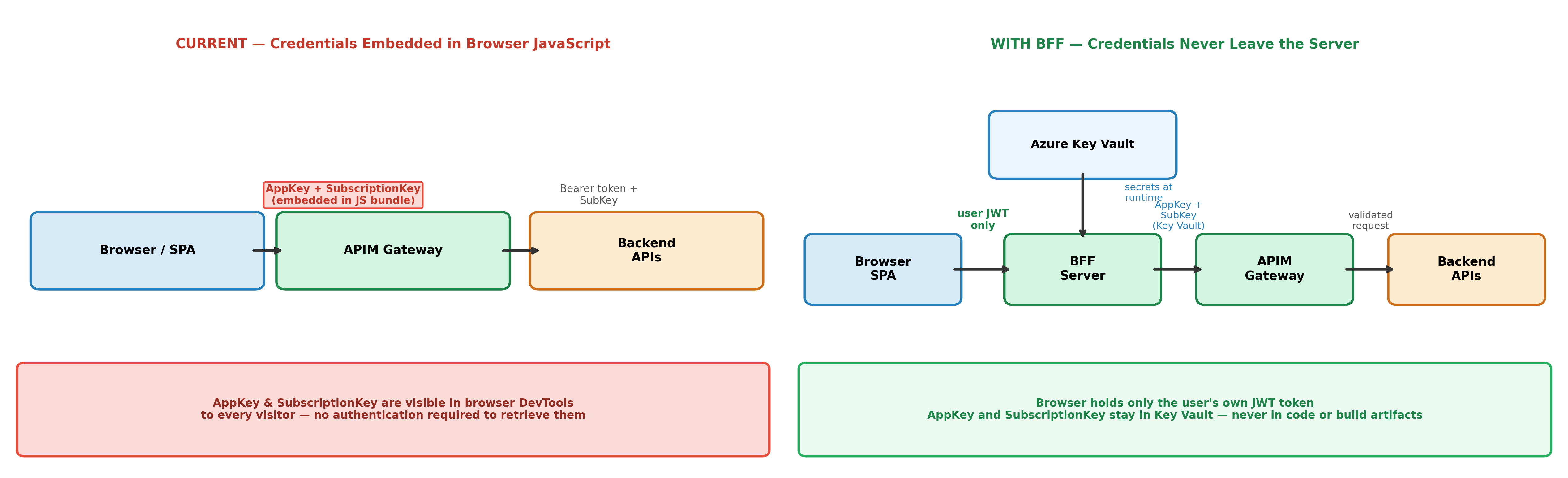}}
\caption{Current architecture (left) versus the Backend for Frontend pattern (right). In the current architecture, AppKey and SubscriptionKey are embedded in the browser JavaScript bundle. The BFF pattern moves all service credentials to a server-side proxy that holds them in Azure Key Vault, so the browser holds only the user's own JWT.}
\end{figure}

The BFF stores \passthrough{\lstinline!AppKey!} and \passthrough{\lstinline!SubscriptionKey!} in Azure Key Vault or App Service configuration, never in code or environment files that reach the build artifact. It receives the user's Azure AD JWT, validates it, and makes downstream calls using its own server-side credentials. The browser never holds anything other than the user's own access token.

Implementation options by effort level:

{\def\LTcaptype{none} 
\begin{table}[!ht]
\centering
\caption{Backend-for-frontend implementation options for removing credentials from the browser, by effort level.}\label{tab:reco}
\adjustbox{max width=\textwidth}{%
\begin{tabular}{@{}
l
  l
  l@{}}
\toprule\noalign{}
Option
 & Effort
 & Notes
 \\
\midrule\noalign{}

\bottomrule\noalign{}

\textbf{Azure Function proxy} & Low & Lightweight, scales automatically, fits alongside existing Azure infrastructure \\
\textbf{Node.js / Express proxy} & Low-Medium & Familiar to Angular teams, can share TypeScript types \\
\textbf{APIM validate-jwt policy} & Low & APIM validates user JWT and rejects requests without one, SubscriptionKey removed from browser entirely \\
\end{tabular}%
}
\end{table}

}

\textbf{Correct the service principal scope.} Replace Application permissions with Delegated permissions where possible. An application acting on behalf of a user should see only what that user is authorized to see, not every user's data. This is the configuration that converted a credential exposure into an account takeover chain in both cases documented above.

\textbf{Restrict MSAL.js token caching to in-memory storage.} The MSAL.js default caches tokens in \passthrough{\lstinline!localStorage!}, which is accessible to any JavaScript executing on the page, including XSS payloads \cite{ref5}. Configuring \passthrough{\lstinline!cacheLocation: "memory"!} limits token extraction risk at the cost of tokens being lost on page refresh.

\textbf{Audit CORS policy on APIM.} Development CORS policies allowing wildcard origins (\passthrough{\lstinline!*!}) with credentials enabled are commonly left open in production. With exposed credentials and permissive CORS, a malicious third-party site can make authenticated API calls using a visiting user's session. Lock CORS to specific, explicitly allowed origins.

\textbf{For non-Azure stacks:} The same architectural principle applies across providers. AWS applications should move Cognito client secrets and API Gateway keys to Lambda or ECS task roles using IAM instance profiles, never into the React bundle. GCP applications should replace Firebase service account keys in the browser with Firebase App Check and server-side token exchange. SaaS credentials (Twilio, Stripe, SendGrid) should be proxied through a server-side endpoint in all cases, no SaaS secret key belongs in client-side JavaScript regardless of platform.

\subsection{Ongoing, Process}\label{ongoing-process}

\textbf{Add artifact scanning to the CI/CD pipeline before deployment.} This closes part of the build-time injection gap by scanning the build artifact before it reaches production. It catches pipeline-substituted secrets that materialized in \passthrough{\lstinline!dist/!} but does not cover runtime-fetched configurations. The scanning step should be inserted after the build stage and before the deployment stage, scanning the compiled output directory rather than source files:

\begin{figure}[!htbp]
\centering
\pandocbounded{\includegraphics[keepaspectratio,alt={Recommended CI/CD pipeline with artifact scan stage inserted between Build and Deploy. The artifact scan targets the compiled output directory (e.g.~dist/), not source files. Secrets introduced by pipeline variable substitution are caught before reaching production.},keepaspectratio,width=\textwidth,keepaspectratio]{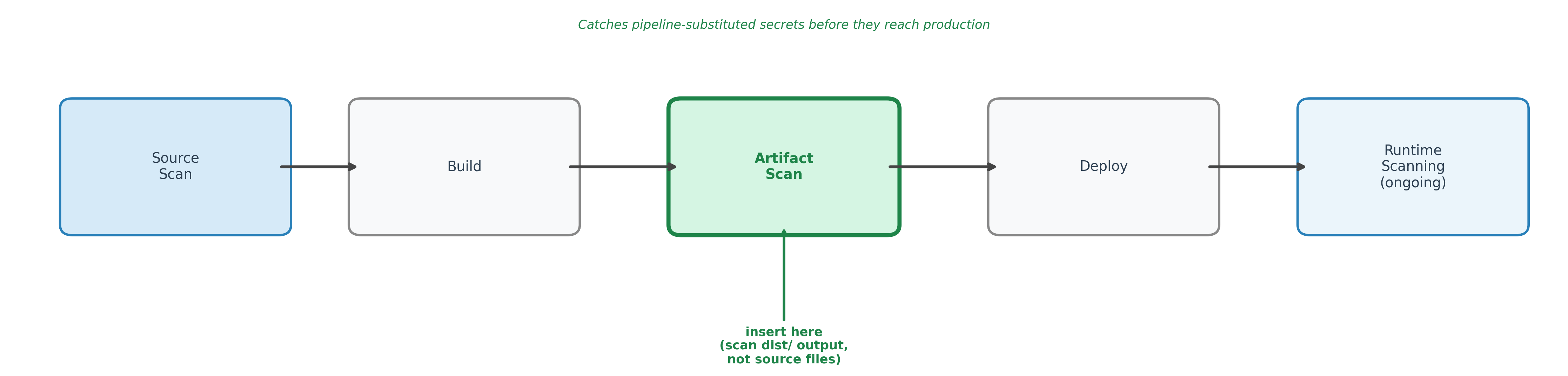}}
\caption{Recommended CI/CD pipeline with artifact scan stage inserted between Build and Deploy. The artifact scan targets the compiled output directory (e.g.~dist/), not source files. Secrets introduced by pipeline variable substitution are caught before reaching production.}
\end{figure}

The artifact scan step targets the compiled output directory (e.g.~\passthrough{\lstinline!dist/!}), not source files. These are different scan targets. On secrets found, the pipeline should fail and block deployment. Any secret scanning tool with filesystem scanning capability is applicable.

\textbf{Monitor for abuse of already-exposed credentials.} Rotation confirms exposure is closed. It does not confirm the credentials were not already used. After rotation:

\begin{itemize}
\tightlist
\item
  Filter APIM Analytics by the rotated subscription key for historical usage anomalies
\item
  Search Azure AD sign-in logs for \passthrough{\lstinline!client\_credentials!} grants originating from browser IP ranges. This flow is designed for server-to-server communication and should not be initiated from end-user browsers
\item
  Alert on any usage of a rotated key post-rotation. This is an active attacker with a cached credential
\item
  Azure Sentinel can correlate APIM logs with Azure AD sign-in anomalies for sustained monitoring
\end{itemize}

\textbf{Eliminate server-side secrets with Azure Managed Identity.} For the BFF or backend layer, Managed Identity removes the need to store or rotate \passthrough{\lstinline!AppKey!} entirely \cite{ref13}:

\begin{lstlisting}
Azure App Service (BFF)
  - Managed Identity enabled
  - No credentials in code, config, or pipeline
  - Azure handles authentication transparently
  - BFF calls APIM / Key Vault / Storage with zero stored secrets
\end{lstlisting}

Combined with Key Vault references in App Service configuration for any remaining secrets, this eliminates the credential rotation problem for the server-side layer.

\textbf{Add periodic runtime scanning of live deployed applications.} Artifact scanning before deployment does not cover runtime-fetched configurations, lazy-loaded chunks served dynamically from CDN, or credentials embedded in third-party scripts that never touch the build pipeline. Periodic scanning of the live application closes this remaining gap. Tools designed for this layer passively monitor live HTTP traffic (JavaScript bundles, HTML source, JSON and XML responses, request headers) and flag credentials as they flow, without outbound verification calls. SecretSifter (the open-source Burp Suite extension evaluated in §8.5) is one such implementation; comparable runtime-aware tooling is appropriate provided it operates on served content rather than source repositories.

\begin{center}\rule{0.5\linewidth}{0.5pt}\end{center}

\section{Limitations}\label{limitations}

The exploitation chains documented in this paper are specific to Azure Active Directory and Azure API Management. The structural paths by which secrets reach production (build-time injection, CI/CD variable substitution, runtime configuration fetching, and scanner suppression) are platform-agnostic and appear across AWS, GCP, and SaaS stacks, as noted in Section 5.3.1. However, the specific credential types, token endpoint formats, and remediation guidance in Sections 2 and 9 are Azure-specific and should be adapted to the relevant platform.

The empirical figures cited in Section 1 are drawn from a single authorized security engagement covering approximately 2,000 enterprise web application assets within one organization, of which 113 contained credentials in served content. The sample is not randomly selected across organizations. Clopper-Pearson exact 95\% confidence intervals are reported alongside each figure: 5.65\% (CI: 4.7\%--6.8\%) for at least one live credential across the 2,000-asset scope, and 55.8\% (CI: 46.1\%--65.1\%) for a complete Azure AD credential set among the 113 credential-bearing applications. Because the sample is drawn from a single organization's portfolio (predominantly Angular SPAs with a shared build pipeline) these figures should not be interpreted as a cross-industry population-level prevalence estimate. A multi-organization prospectively designed study would be required to produce generalizable estimates. Independent industry-scale measurement provides external corroboration of the underlying exposure pattern: Intruder's December 2025 sweep \cite{ref28} reported approximately 42,000 exposed tokens across roughly 5 million applications using regex-driven static detection, a finding consistent with the prevalence direction reported here while differing in scale and detection methodology.

Section 8.5 presents a comparison evaluation of nine production scanners against the locked Ground Truth GT-194, 194 unique secret-grade credentials extracted across 113 enterprise applications by Claude Opus 4.7 (reported separately as the LLM-assisted ground-truth reference comparator, not as an evaluated detector), its LLM-extracted candidates cross-validated by GPT-5.5 (Brennan-Prediger κ = 0.676 over the 247 LLM-extracted candidates), and locked at the field-deployment register on 8 May 2026. The evaluation has five documented limitations.

\textbf{(1) Single-organization corpus and author-developer role.} The application pool is drawn from one engagement within one organization (predominantly Angular SPAs with a shared build pipeline). The first author (Gorijala) is also the developer of SecretSifter, one of the evaluated scanners. The headline recall numbers in Table~\ref{tab:recall} should therefore be read as scanner performance on this specific Azure-heavy corpus, not as cross-organizational population estimates. The structural mitigations applied (independent LLM Ground Truth extraction, cross-vendor classification, exact-value substring matching with no per-tool tuning of the comparison protocol, the same 6-rule overlay applied identically to all rule-extensible tools) prevent the comparison from being engineered toward a specific tool, but they do not eliminate the corpus-specificity of the absolute numbers. A multi-organization replication on an independently constructed held-out benchmark would be required to claim cross-industry generalization, and is committed to as a direct follow-on study.

\textbf{(2) Sample frame.} The 113-application benchmark subset is drawn from the same engagement that produced the Ground Truth, not from an independently sampled population. Recall and per-class results should be interpreted as measurements of scanner performance against GT-194, not as cross-organizational performance estimates. The ground-truth construction methodology (independent LLM extraction with cross-vendor validation) is the contribution that generalizes. The recall figures are anchored to this corpus.

\textbf{(3) Tightened-configuration scope.} The 6-rule overlay applied to the rule-extensible tools (TruffleHog, SecretFinder, Titus, JSluice) was sized to cover the most prevalent credential shapes in this corpus and applied identically across all four tools (Appendix A). A more aggressive tuning frontier, full vendor-rule libraries, customer-specific patterns, or per-tool optimization, may further close the static-tool gap. The point of the comparison is not that 6 rules are enough; it is that even a corpus-targeted overlay applied uniformly leaves a residual gap of 41 percentage points to a runtime-aware scanner, on this corpus.

\textbf{(4) Ground-truth reference limitations.} The LLM-assisted ground-truth reference is itself a model with its own biases. Claude Opus did not surface 5 of 37 CryptoJS-AES blobs and 18 of 153 Azure AD App\_IDs in the chain-context tier, demonstrating that the reference model has structural blind spots that curated detection rules complement. The CryptoJS blobs the reference comparator missed were recovered by SecretSifter and are therefore not part of the tool-agnostic 27; that blind spot instead comprises credentials surfaced only by manual analysis, so the gap is a property of the underlying credential corpus rather than an LLM-vs-static comparison artefact.

\textbf{(5) Per-credential vs per-application generalization.} GT-194 is a per-credential ground truth; the chain-completion finding in §8.5.7 is a per-application corpus property derived from 86 secret-exposed applications. The 73.3\% (63/86) figure is descriptive of this engagement, not inferential of cross-industry chain-completion rates. A multi-organization replication is required for inferential claims about full-chain reconstruction rates. Relatedly, credentials are clustered within applications (multiple GT-194 credentials often originate from the same application bundle), so the per-credential observations that the McNemar and Holm-Bonferroni analyses (§8.5.10) treat as units are not fully independent deployment events. The significance results should therefore be read as pairwise detection-difference tests over the credential set, not as inferences over independent applications; a cluster-aware or per-application analysis is left to multi-organization replication.

\begin{center}\rule{0.5\linewidth}{0.5pt}\end{center}

\section{Future Work}\label{future-work}

\textbf{Multi-organization prevalence study.} The 5.65\% credential exposure rate reported in this paper is drawn from a single organization's application portfolio. A prospectively designed study across multiple organizations (stratified by industry, stack, and deployment maturity) is required to produce a generalizable prevalence estimate. A multi-organization replication study, comparable in scope to the 2,000-application engagement reported here, is planned as a direct follow-on.

\textbf{Cross-organization precision benchmark.} §8.5.5 reports direct precision, recall, and F1 for the nine production scanners against the GT-194 corpus (with Claude Opus reported separately as the ground-truth reference comparator). Replicating the same precision measurement on a held-out, multi-organization production corpus, with full TP / FP classification of every emitted finding by an independent validator, would test whether the precision ordering reported here transfers across stacks. The qualitative ordering (parser-based scanners produce few false positives, regex-based scanners produce many, SecretSifter's noise-suppression layer materially differentiates it from comparable static tools) is robust under any reasonable re-measurement, but the exact precision and F1 numbers per scanner would tighten.

\textbf{Multi-organization held-out benchmark.} GT-194 is constructed from a single engagement. Replicating the same construction methodology on a held-out benchmark drawn from multiple unrelated organizations would test whether the recall numbers reported here transfer across stacks, build pipelines, and credential conventions. The construction methodology (Claude Opus extraction + GPT-5.5 cross-vendor validation) is fully described in §8.5.1 and is reproducible by any practitioner with API access to both vendors.

\textbf{Runtime-fetched configuration coverage.} GT-194 evaluates scanners against downloadable JavaScript bundles only, the surface for which static-file scanners can be applied uniformly. Modern SPAs load credentials via runtime-fetched configuration endpoints (\passthrough{\lstinline!/config.json!}, \passthrough{\lstinline!/api/settings!}) and HTML SSR state blobs that are never present in the initial bundle. Extending the benchmark methodology to runtime-fetched and SSR-injected credentials is the natural next step and would close the gap between the bundle-scanning protocol described here and the production-traffic surface that real attackers actually inspect.

\textbf{Automated exploitation chain validation.} The current study manually confirmed exploitability for both documented chains. Automated scope-aware validation, checking whether exposed Azure AD credentials have \passthrough{\lstinline!client\_credentials!} grant enabled, or whether exposed APIM keys map to APIs with sensitive data, would allow prioritization of findings at scale without manual verification per application.

\textbf{Longitudinal remediation tracking.} Both applications documented in this paper remediated within the assessment window. Whether organizations with incidentally discovered runtime credential exposure (found via bug bounty or passive scanning) remediate at comparable rates is unknown. A longitudinal study tracking time-to-remediation across disclosure channels would inform remediation guidance.

\begin{center}\rule{0.5\linewidth}{0.5pt}\end{center}

\section{Responsible Disclosure}\label{responsible-disclosure}

Both exploitation chains documented in this paper were identified during authorized security assessments conducted within the scope of formal engagements. In both cases:

\begin{itemize}
\tightlist
\item
  Findings were reported immediately to the affected organization upon identification
\item
  Proof-of-concept evidence was limited to what was necessary to confirm exploitability, no bulk data extraction occurred
\item
  Remediation was implemented by the affected organization
\item
  Findings were verified through retest to confirm remediation before this publication
\end{itemize}

No user data was retained beyond the assessment period. No credentials were used outside the scope of confirming exploitability. Organization names and application identifiers have been anonymized throughout this paper.

\begin{center}\rule{0.5\linewidth}{0.5pt}\end{center}

\section{Conclusion}\label{conclusion}

The shift-left investment the security industry has made is real, justified, and valuable. Catching secrets before they reach production is always preferable to finding them after. GitLeaks, TruffleHog, GitHub Advanced Security, and the SAST tools deployed in enterprise CI/CD pipelines do exactly what they are designed to do.

But secrets are still reaching production, not because these tools fail, but because the path from development to deployment has branches that no pre-deployment scanner covers. Build-time environment injection, CI/CD pipeline variable substitution, and runtime configuration fetching each produce secrets in production that never existed in any repository at any point. The tools designed to prevent credential exposure scan a layer that those secrets never passed through.

The exploitation chains documented above are not theoretical edge cases. They are patterns found repeatedly in real production applications serving real users, applications that had passed every security review and had shift-left tooling in place. None of those tools provided automated visibility into what each application served, or automated detection when credentials appeared in delivered JavaScript.

The shift-right tooling gap is now quantified, and includes a tool-agnostic blind spot. On the locked Ground Truth GT-194, 13.9\% of credentials are missed by every one of the nine production scanners evaluated and are surfaced only by manual analysis. The CryptoJS encrypted-configuration class is invisible to every static scanner by design and is recovered only by runtime-aware and LLM-based detection. The combined coverage of the nine production scanners together with the ground-truth reference comparator plateaus at 86.1\%, leaving that 13.9\% as a structural property of the credential surface itself rather than a tooling deficit that more rules can close. Beyond detection, the chain-completion finding (73.3\% of affected applications co-locate the full Azure AD token-mint chain client-side) shows that the credentials reachable in served JavaScript are operationally weaponizable from browser-visible code alone. Industry-scale measurement corroborates this gap: Intruder's December 2025 sweep, an industry vendor report \cite{ref28}, identified approximately 42,000 exposed tokens across roughly 5 million applications using regex-driven static detection alone, a lower-bound observation that the pattern documented here is widespread at industrial scale. The shift-left layer is covered. The runtime layer is not. That asymmetry, and the additional 13.9\% blind spot that no evaluated production scanner closes on this corpus, is what is being exploited.

Closing it requires the same community effort that built shift-left tooling into what it is today, practitioners documenting the gap, security teams demanding runtime scanning in their programs, and tool authors building for the layer that has been ignored. Bug bounty researchers finding runtime credentials should escalate beyond the credential itself and document the full exploitation chain. Penetration testers should treat runtime JavaScript as a first-class target surface, not an afterthought. Security teams should ask their tooling vendors a direct question: does your scanner cover what the application serves, or only what developers commit?

The patterns in this paper will not stop appearing until the runtime layer gets the same tooling investment the repository layer already has.

\begin{center}\rule{0.5\linewidth}{0.5pt}\end{center}

\appendices
\section{Tightened-Configuration Rule Set}\label{appendix-c-tightened-configuration-rule-set}

The four rule-extensible scanners evaluated in §8.5 (TruffleHog, SecretFinder, Titus, JSluice) were each run with the same 6-rule overlay. The rules cover the credential shapes most prevalent in GT-194. The full rule set is reproduced below verbatim. The same overlay was applied identically to all four tools; no per-tool tuning was performed.

\begin{lstlisting}
- id: custom.azure.client_id
  name: Azure AD Client ID (UUID v4)
  pattern: '\b[0-9a-f]{8}-[0-9a-f]{4}-[0-9a-f]{4}-[0-9a-f]{4}-[0-9a-f]{12}\b'

- id: custom.azure.ad_secret
  name: Azure AD Client Secret (tilde-format, 34-40 char)
  pattern: '\b[A-Za-z0-9_.\-]{1,16}~[A-Za-z0-9_.\-~]{16,40}\b'

- id: custom.azure.apim_key
  name: Azure APIM Subscription Key (32-hex)
  pattern: '\b[0-9a-f]{32}\b'

- id: custom.jwt.token
  name: JSON Web Token (eyJ-prefix three-segment)
  pattern: '\beyJ[A-Za-z0-9_\-]{10,}\.[A-Za-z0-9._\-]{10,}\.[A-Za-z0-9._\-]+'

- id: custom.azure.app_insights_ikey
  name: App Insights Instrumentation Key
  pattern: 'InstrumentationKey=[0-9a-f]{8}-[0-9a-f]{4}-[0-9a-f]{4}-[0-9a-f]{4}-[0-9a-f]{12}'

- id: custom.cryptojs.salted
  name: CryptoJS U2FsdGVkX1-prefix encrypted blob
  pattern: 'U2FsdGVkX1[A-Za-z0-9+/=]{16,}'
\end{lstlisting}

These six rules were selected from the credential shapes most prevalent in GT-194, derived without consulting any tool's output. They are not a comprehensive coverage set; they are a minimal corpus-targeted overlay applied identically to \textbf{all four} rule-extensible tools (TruffleHog, SecretFinder, Titus, JSluice) so that the comparison reflects scanner-architecture differences rather than rule-library differences. JSluice was run with the same six patterns translated to the JSluice user-pattern JSON schema; the built-in JSluice secret rules silently produce zero output on minified webpack bundles in this corpus, and the user-pattern overlay is what allows JSluice to surface the key-value extractions that drive its reported recall. JSMiner exposes no user-rule API in the Burp extension distribution, so it was reimplemented in Python as a faithful clone of the published Trustwave SpiderLabs detection logic (21 keyword anchors plus Shannon-entropy scoring with the same Firm/Tentative classification thresholds) and run on the GT-194 corpus. Nuclei was run in the URL-mode JavaScript-spider configuration matching the Intruder December 2025 industry-scale measurement \cite{ref28}: bundles served via a local Python HTTP server on port 9101, scanned with the secret-scanning template categories \passthrough{\lstinline!http/exposures/tokens!}, \passthrough{\lstinline!http/exposures/apis!}, and \passthrough{\lstinline!http/exposures/keys!} plus the \passthrough{\lstinline!secret!}/\passthrough{\lstinline!token!}/\passthrough{\lstinline!key!}/\passthrough{\lstinline!exposure!} template tags. The file-mode Nuclei configuration produces zero findings on raw .js files because Nuclei's secret-scanning template library is HTTP-response-oriented rather than file-content-oriented, a methodology mismatch documented here for reproducibility. Cariddi was run at its built-in default secret-detection configuration without any rule overlay (its primary mode is web-crawling and endpoint enumeration, with secondary secret detection). SecretSifter was run at its built-in configuration with the field-deployment scan output exported via the SecretSifter REST API.

\begin{center}\rule{0.5\linewidth}{0.5pt}\end{center}

\emph{Author contributions (CRediT): \textbf{Gorijala}, conceptualization, methodology, software (SecretSifter), investigation, validation, formal analysis, data curation, writing of original draft, writing of review and editing, project administration.}

\emph{Declaration of generative AI use: This work used AI tools in two ways. As research instruments, Claude Opus 4.7 (Anthropic) built the GT-194 ground-truth benchmark and GPT-5.5 (OpenAI) independently validated its LLM-extracted candidates, with the manual-only additions verified by analyst review; Section 8.5 documents this and cites both systems as \cite{ref22} and \cite{ref23}. Generative AI also assisted with drafting and language editing of the manuscript. The author checked every statistic, citation, and technical claim, revised the text, and is fully responsible for its accuracy, originality, and integrity. No AI system is an author. No AI tool produced research data, results, or analysis beyond the benchmark construction described in Section 8.5.}

\emph{Conflict of interest disclosure: The author (Gorijala) is the developer of SecretSifter, which is referenced in Section 9.3 as a tool for runtime scanning and evaluated as one of the nine production scanners in Section 8.5. The author's role as developer is mitigated structurally in the §8.5 evaluation: the Ground Truth (GT-194) was constructed by independent Claude Opus 4.7 (Anthropic) extraction and manual analyst review, and independently validated by GPT-5.5 (OpenAI), so no scanner (including SecretSifter) defines its own evaluation set; the union also contains credentials SecretSifter did not detect, making the benchmark strictly harder for it rather than easier. The SecretSifter edition evaluated here (the Burp Suite extension) is open source. The exploitation chains, structural gap analysis, and prevalence figures in this paper are independent of that tool and predate its development.}

\EOD

\end{document}